\documentclass{article}
\usepackage[T2A]{fontenc}
\usepackage[utf8]{inputenc}
\usepackage[english]{babel}
\usepackage{amsmath}
\usepackage{amsfonts}
\usepackage{amssymb}
\usepackage{wasysym}
\usepackage{amsthm}
\usepackage{tikz}
\usepackage{tikz-cd}
\usetikzlibrary{decorations.markings}
\usetikzlibrary{positioning}
\usepackage{array}
\usepackage{ytableau}
\usepackage{longtable}
\usepackage{empheq}
\usepackage{multicol}
\usepackage{mathrsfs}
\usepackage{amssymb}
\usepackage{diagbox}
\usepackage{physics}
\usepackage{pb-diagram}
\usepackage{setspace}
\usepackage{faktor}
\usepackage{rotating}
\usepackage[hidelinks,colorlinks=true,unicode]{hyperref}
\hypersetup{linkcolor=blue, citecolor=blue, filecolor=blue, urlcolor=blue}
\usepackage{url}
\usepackage[left=2cm,right=2cm,top=2cm,bottom=2cm,bindingoffset=0cm]{geometry}

\usepackage[natbib=true, backend = bibtex, style=numeric-comp, sorting=none]{biblatex}
\def\be{\begin{eqnarray}}
\def\ee{\end{eqnarray}}

\begin{document}
 
\title{{\Large {\bf Polynomial representations of classical Lie algebras and flag varieties
}}
\author{
{\bf A.~Morozov $^{a,b,c}$}\thanks{morozov.itep@mail.ru},
{\bf M.~Reva $^{a,c}$}\thanks{reva.ma@phystech.edu},
{\bf N.~Tselousov $^{a,c}$}\thanks{tselousov.ns@phystech.edu},
{\bf Y.~Zenkevich $^{c,d,e,f,g}$}\thanks{yegor.zenkevich@gmail.com} \date{ }}
}
\maketitle

\vspace{-5cm}
 
\begin{center}
    \hfill MIPT/TH-02/22\\
    \hfill IITP/TH-02/22\\
    \hfill ITEP/TH-03/22
\end{center}
 
\vspace{2.5cm}

\begin{center}

\begin{small}
$^a$ {\it MIPT, Dolgoprudny, 141701, Russia}\\
$^b$ {\it IITP, Moscow 127994, Russia}\\
$^c$ {\it ITEP, Moscow 117218, Russia}\\
$^d$ {\it ITMP, Moscow 119991, Russia}\\
$^e$ {\it SISSA, Trieste, Italy}\\
$^f$ {\it INFN, Sezione di Trieste}\\
$^g$ {\it IGAP, Trieste, Italy}\\
\end{small}
 
\end{center}
 
\vspace{1cm}

\begin{abstract}
Recently we have started a program to describe
the action of Lie algebras associated with Dynkin-type diagrams on generic Verma modules in terms of polynomial vector fields.
In this paper we explain that the results for
the classical ABCD series of Lie algebras coincide with
the more conventional approach,
based on the knowledge of the entire algebra,
not only the simple roots.
We apply the coset description,
starting with a large representation and then reducing it with the help of the algebra, commuting with the original one. The irreducible representations are then obtained
by gauge fixing this residual symmetry.
\end{abstract}

\bigskip

\bigskip

\section{Introduction}
\label{sec:intro}

This paper is a step in the new program, targeted to systematically describe
polynomial representations of arbitrary Dynkin diagrams and their futher
generalization to elliptic and DIM algebras.
The initial piece, concerning classical Lie algebras ($A$, $B$, $C$, $D$ series),
has many overlaps with traditional presentations (see \cite{Dobrev1,Dobrev2,Dobrev3,Dobrev4, KrizkaSomberg1,KrizkaSomberg2, KrizkaSomberg4, Shchepochkina, Leites1, Leites2, Leites3,Leites4} and reference therein),
Kac-Moody case will lead to the free-field description \cite{GMMOS,FeFr, KrizkaSomberg3},
further generalizations are related to other more recent results in representation theory.
Our purpose is to build a uniform and universal construction in the most elementary way,
using no detailed knowledge about particular algebras
understandable to a broad audience.
In current paper we explain how the results, obtained in this way
for classical algebras in \cite{MRTZ1}, are related to the language of
coset theories, gauge transformations and flag varieties
which is perhaps more familiar to some readers.

We begin with the conventional description of classical Lie algebra $\mathfrak{g}$ as an  algebra
of $N\times N$ matrices. Consider the
matrix generators 
$E_{i,j}$, $i,j=1,\ldots,N$ with standard commutation relations
\begin{equation}
\label{gl generators}
    [E_{i,j}, E_{k,l}] = \delta_{j,k} \, E_{i,l} - \delta_{i,l} \, E_{k,j}.
\end{equation}
where $N$ is a dimension of fundamental/defining representation (see Table \ref{Dimensions of fundamental rep}). 

\begin{table}[h]
\begin{doublespace}
    \centering
    \begin{tabular}{|c|c|c|c|c|}
    \hline
         Root system &  $\text{A}_n$ & $\text{B}_n$ & $\text{C}_n$ & $\text{D}_n$ \\
         \hline
         Lie algebra &  $\mathfrak{sl}_{n+1}$ & $\mathfrak{so}_{2n+1}$ & $\mathfrak{sp}_{2n}$ & $\mathfrak{so}_{2n}$ \\
         \hline
         $N$ & $n+1$ & $2n+1$ & $2n$ & $2n$ \\
         \hline
    \end{tabular}
\end{doublespace}
\caption{\footnotesize{The dimension $N$ of the fundamental/defining representation for classical Lie algebras. $n$ is a rank of the algebra.}}
\label{Dimensions of fundamental rep}
\end{table}

We consider a particular representation of elementary matrices $E_{i,j}$:
\begin{equation}
\label{E repr}
    \rho\left( E_{i,j} \right) = \sum_{a=1}^{n} x_{i}^{a} \frac{\partial}{\partial x^{a}_{j}}
\end{equation}
made from rectangular $n \times N$ matrix $x_{i}^{a}$, where $i=1,\ldots,N$, $a = 1,\ldots,n$. The most interesting feature of this representation is that it allows to describe arbitrary representation of any, arbitrary large "spin" in terms of vector fields, acting on just $n \times N$ matrices $x_{i}^{a}$:
different representations are actually formed by different powers of $x_{i}^{a}$, this is why they are called {\it polynomial}. For $n < \rank \mathfrak{g}$ this representation is actually too small,
namely, does not contain all irreducible representation.
For $n= \rank \mathfrak{g}$ it does contain all irreducible representations, moreover there are several copies of each one. Turns out that there is a whole bunch of operators, which commute with $\rho \left( E_{ij} \right)$ and act between several copies of the irreducible representations. There are two usual choices for this complementary set of operators: one can consider Casimirs, which are {\it poly}vectors, but scalars (do not have "open" indices) or, alternatively, one can take an entire {\it dual}\footnote{By the term "dual algebra" we {\it do not} mean the ordinary definition of dual space $\mathfrak{g}^{*}$ to the Lie algebra $\mathfrak{g}$. Hope, it does not cause a confusion.} algebra of matrix-valued vector fields $\tilde{\rho} \left(\tilde{\mathfrak{g}}\right)$ .
These operators form a Lie algebra $\tilde{ \mathfrak{g}}$, that in some sence is dual to the initial algebra $\mathfrak{g}$. \\

Our aim in this paper is to obtain {\it all irreducible representations of classical Lie algebras in terms of polynomial vector fields}. We do it considering representation $n= \rank \mathfrak{g}$ and picking out one particular copy of each irreducible representation exploiting the commutativity relation:
\begin{equation}
\label{Algebra and dual commute}
    \left[ \rho(\mathfrak{g}), \tilde{\rho} (\tilde{\mathfrak{g}}) \right] = 0
\end{equation}
where $\tilde{\rho}\left( \tilde{\mathfrak{g}} \right)$ denotes operators of the dual/commuting algebra $\tilde{\mathfrak{g}}$. In other words, we consider commuting operators $\tilde{\rho}\left( \mathfrak{g} \right)$ as a "symmetry of the system" and pick out "physical degrees of freedom" by conducting a proper procedure of a reduction over the "symmetry". This point of view make the story close to the well-known theory of Hamiltonian reduction.
\\


Information about this dual algebra $\tilde{\mathfrak{g}}$ is given in Table~\ref{Commuting algebras generators}. The appearance of the dual commuting algebra seems to be a general phenomenon, when the initial representation contains ``too many'' variables. This is a particular case of Howe duality~\cite{Howe1, Howe2}.

For general simple Lie algebra, it is an open question how to write down the generators of the dual algebra. However, the classical series of Lie algebras is much simpler due to the presence of additional structures, such as invariant bi-linear forms.

Both representations $\rho(\mathfrak{g}), \tilde{\rho} (\tilde{\mathfrak{g}})$ act on the space of polynomials $\mathbb{C}[x_{i}^{a}]$. The space $\mathbb{C}[x_{i}^{a}]$ decomposes into a direct sum of tensor products of irreducible representations of $\mathfrak{g}$ and $\tilde{\mathfrak{g}}$ which are in one-to one correspondence:
\begin{equation}
\label{Polynomial space splits}
    \mathbb{C}[x_{i}^{a}] = \bigoplus_{\lambda} R_{\lambda}^{\mathfrak{g}} \otimes R_{\sigma(\lambda)}^{\tilde{\mathfrak{g}}}.
\end{equation}

Here $\lambda$ labels both highest weights of $\mathfrak{g}$ while $\sigma(\lambda)$ is the highest weight of $\tilde{\mathfrak{g}}$ corresponding to $\lambda$. Eq.~\eqref{Polynomial space splits} is understood in the following sense: for every $\lambda$ the space $\mathbb{C}[x_{i}^{a}]$ contains several copies of representation $R_{\lambda}^{\mathfrak{g}}$, which can be labelled by the states of the representation $R_{\tilde{\lambda}}^{\tilde{\mathfrak{g}}}$. Root generators of $\tilde{\mathfrak{g}}$ act between different copies of $R_{\lambda}^{\mathfrak{g}}$. Similar statement is true if one exchanges $\mathfrak{g}$ and $\tilde{\mathfrak{g}}$.

The most interesting feature of the representation \eqref{E repr} is that
it allows one to describe \emph{arbitrary} irreducible representation
in terms of vector fields, acting on the space of polynomials:
different representations are obtained from different monomials in $x_{i}^{a}$. The sum \eqref{Polynomial space splits} contains all highest weights $\lambda$.

We aim to describe all irreducible highest weight representations $R_{\lambda}^{\mathfrak{g}}$ of $\mathfrak{g}$ in terms of vector fields in $\mathbb{C} \left[ x_{i}^{a} \right]$. We do it by choosing for every $\lambda$ one particular copy of $R_{\lambda}^{\mathfrak{g}}$ in $\mathbb{C} \left[ x_{i}^{a} \right]$ and make suitable projection/reduction on it. After the reduction the number of remaining variables is equal to the number of positive roots $| \Delta_{+}^{\mathfrak{g}}|$ of algebra $\mathfrak{g}$. This kind of representations are actually the representation on the corresponding flag variety. We explain this aspect in detail for each series.

In particular we make the reduction on the copy of $R_{\lambda}^{\mathfrak{g}}$ that is the space of all highest vectors of weight $\tilde{\lambda}$ of $\tilde{\mathfrak{g}}$. Schematically, each element $\ket{\lambda}$ of this space obey the following equations:
\begin{align}
\label{Space of highest vectors}
\begin{aligned}
    \mathbf{e}_{\alpha} \ket{\lambda} &= 0, &\hspace{10mm} &\text{for} \ \alpha \in \Delta_{+}^{\tilde{\mathfrak{g}}}  \\
    \mathbf{h}_i \ket{\lambda} &= \tilde{\lambda}_i \ket{\lambda}, &\hspace{10mm} &\text{for} \ i = 1, \ldots, \rank \tilde{\mathfrak{g}}
\end{aligned}
\end{align}

The space of highest vectors \eqref{Space of highest vectors} forms an irreducible representation $R_{\lambda}^{\mathfrak{g}}$ and we consider {\it restriction} on this space and use equations \eqref{Space of highest vectors} to express some variables and derivatives through the others. As a result, we obtain an irreducible representation in terms of vector fields on $| \Delta_{+}^{\mathfrak{g}}|$ variables with some free parameters. "Fixing the gauge" we set this additional parameters to 0 or 1 to get the simplest form of the formulas. 
Let us summarize the reduction process as the following algorithm:
\begin{enumerate}
    \item Consider the fundamental/defining representation of classical Lie algebras $\mathfrak{g}$ as $N \times N$ matrices.
    \item Find a basis of simple root generators (Chevalley basis).
    \item Construct $\rho \left( \mathfrak{g} \right)$ representation that acts on the space of polynomials $\mathbb{C} \left[ x_{i}^{a} \right]$ by mapping elementary matrices $E_{i,j}$ to differential operators \eqref{E repr}. 
    \item Find the algebra of commuting generators $\tilde{\rho} (\tilde{\mathfrak{g}})$ satisfying \eqref{Algebra and dual commute}.
    \item  Consider equations of type \eqref{Space of highest vectors}  describing the space of highest vectors of algebra $\tilde{\mathfrak{g}}$ and forming an irreducible representation of $\mathfrak{g}$. The number of equations is $|\Delta_{+}^{\tilde{\mathfrak{g}}}| + \rank{\tilde{\mathfrak{g}}}$.
    \item Choose $|\Delta_+^{\mathfrak{g}}|$ variables of $x_{i}^{a}$ and corresponding derivatives $\frac{\partial}{\partial x_{i}^{a}}$ that will describe irreducible representations. 
    \item Use equations from step 5 to express remaining variables and corresponding derivatives through the distinguished variables and derivatives from step 6. 
    \item Set the remaining free parameters to constant values.
    \item Substitute the results to the simple roots generators of $\rho(\mathfrak{g})$.
\end{enumerate}
The resulting formulas \eqref{sl(n+1) result},\eqref{so(2n+1) result},\eqref{sp(2n) result},\eqref{so(2n) result} for arbitrary weight $\lambda$
coincide with those, derived from the Dynkin diagrams in \cite{MRTZ1}.
\\
This paper is organized as follows. We start with the simplest example of $A$ series in Sec.\eqref{sec:A} and discuss in detail the subtleties of the reduction/projection procedure. In the next section \eqref{sec:BCD} we provide necessary formulas for B,C,D series and focus on the reduction formulas in section \eqref{sec:BCD reduction}. The formulas for all irreducible representations (simple root generators $\textbf{e}_i,\textbf{f}_i$ and Cartan generators $\textbf{h}_i$) are provided in section \eqref{sec: results}.

\bigskip

\section{Polynomial representation of classical Lie algebras}
\label{sec:A}
\subsection{Starting example of $\text{A}_n$ series}
To describe representations of the algebra $\mathfrak{sl}_{n+1}$ it would be convenient for us to start with a slightly larger algebra
$\mathfrak{gl}_{n+1}$, i.e. in this case the matrix size is $N=n+1$.
This algebra acts on polynomials of $x_{i}^{a}$
and its generators have the following form 
\begin{equation}
  \label{eq:25}
  \rho \left( E_{i,j} \right) = \sum_{a=1}^{n} x_{i}^{a} \frac{\partial}{\partial x_{j}^{a}}
\end{equation}
with $i,j=1,\ldots,n+1$.
One can note that there is a complementary set of operators
\begin{equation}
  \label{eq:25}
  \tilde{\rho}\left( {E}_{a,b} \right) = \sum_{i=1}^{n+1} x_{i}^{a} \frac{\partial}{\partial x_{i}^{b}}
\end{equation}
for $a,b = 1\ldots, n$, that forms dual algebra $\tilde{\mathfrak{g}} = \mathfrak{gl}_{n}$ and commutes with the original one:
\be
\left[\rho \left( E_{i,j} \right), \tilde{\rho}\left( {E}_{a,b} \right) \right] = 0
\ee
This means that the eigenfunctions of any  operator $\tilde{\rho}\left( {E}_{a,b} \right)$ form an invariant
subspace for $\rho \left( E_{i,j} \right)$.
The smallest subspaces of this kind are made from the common eigenfunctions
of as many $\tilde{\rho}\left( {E}_{a,b} \right)$ as possible -- and they have good chances to provide
{\it irreducible} representations of $\rho \left( E_{i,j} \right)$.
Since $\tilde{\rho}\left( {E}_{a,b} \right)$ do not commute, the maximal possibility is to pick up
the common eigenfunctions of their triangular (Borel) sub-algebra,
so that all the non-trivial eigenvalues appear in diagonal (commuting) part, while for the upper-triangular part they are all zeroes:
\begin{equation}
  \label{eq:27}
\boxed{\  \tilde{\rho}\left( {E}_{a,b} \right) \, P(x_{i}^{a})  = \delta_{a,b}\cdot \tilde{\lambda}_a \, P(x_{i}^{a}),\qquad a \leqslant b\ }
\end{equation}
$P(x_{i}^{a})$ is a polynomial in $x_{i}^{a}$, and we parameterize the non-vanishing eigenvalues
$\tilde{\lambda}_a$ in a peculiar way, associated with the true weights $\lambda_a$ of the $\text{A}_n$ representations.
Eq.(\ref{eq:27}) are the analogues of equations \eqref{Space of highest vectors} --
it allows to eliminate the superficial derivatives (momenta). \\
For illustrative purposes we provide explicit example of formula \eqref{Polynomial space splits} for weight $\lambda = 1$ and $n=2$, i.e. representations of unit weight of dual pair $\left( \mathfrak{sl}_3, \mathfrak{gl}_2\right)$:
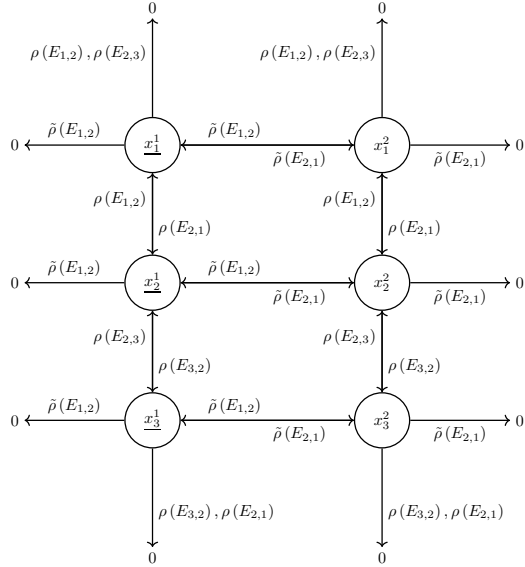
\begin{center}
\resizebox{0.4\textwidth}{!}{%
    \begin{tikzpicture}
        \node (a1) at (-3,3) {$0$};
        \node (a2) at (-3,0) {$0$};
        \node (a3) at (-3,-3) {$0$};

        \node (a4) at (0,6) {$0$};
        \node (a5) at (0,3) [minimum size = 12mm, shape = circle, thick, draw] {$\underline{x_{1}^{1}}$};
        \node (a6) at (0,0) [minimum size = 12mm, shape = circle, thick, draw] {$\underline{x_{2}^{1}}$};
        \node (a7) at (0,-3) [minimum size = 12mm, shape = circle, thick, draw] {$\underline{x_{3}^{1}}$};
        \node (a8) at (0,-6) {$0$};

        \node (a9) at (5,6) {$0$};
        \node (a10) at (5,3) [minimum size = 12mm, shape = circle, thick, draw] {$x_{1}^{2}$};
        \node (a11) at (5,0) [minimum size = 12mm, shape = circle, thick, draw] {$x_{2}^{2}$};
        \node (a12) at (5,-3)[minimum size = 12mm, shape = circle, thick, draw]  {$x_{3}^{2}$};
        \node (a13) at (5,-6) {$0$};
        
        \node (a14) at (8,3) {$0$};
        \node (a15) at (8,0) {$0$};
        \node (a16) at (8,-3) {$0$};

        \draw[thick, ->] (a5)  edge node[above] {$\tilde{\rho}\left( E_{1,2} \right)$} (a1);
        \draw[thick, ->] (a6)  edge node[above] {$\tilde{\rho}\left( E_{1,2} \right)$} (a2);
        \draw[thick, ->] (a7)  edge node[above] {$\tilde{\rho}\left( E_{1,2} \right)$} (a3);
        
        \draw[thick, ->] (a5)  edge node[above left] {$\rho\left( E_{1,2} \right),\rho\left( E_{2,3} \right)$} (a4);
        \draw[thick, ->] (a5)  edge node[below right] {$\rho\left( E_{2,1} \right)$} (a6);
        \draw[thick, ->] (a6)  edge node[above left] {$\rho\left( E_{1,2} \right)$} (a5);
        \draw[thick, ->] (a6)  edge node[below right] {$\rho\left( E_{3,2} \right)$} (a7);
        \draw[thick, ->] (a7)  edge node[above left] {$\rho\left( E_{2,3} \right)$} (a6);
        \draw[thick, ->] (a7)  edge node[below right] {$\rho\left( E_{3,2} \right),\rho\left( E_{2,1} \right)$} (a8);
        
        \draw[thick, ->] (a5)  edge node[below right] {$\tilde{\rho}\left( E_{2,1} \right)$} (a10);
        \draw[thick, ->] (a10) edge node[above left] {$\tilde{\rho}\left( E_{1,2} \right)$} (a5);
        \draw[thick, ->] (a11)  edge node[above left] {$\tilde{\rho}\left( E_{1,2} \right)$} (a6);
        \draw[thick, ->] (a6)  edge node[below right] {$\tilde{\rho}\left( E_{2,1} \right)$} (a11);
        \draw[thick, ->] (a7)  edge node[below right] {$\tilde{\rho}\left( E_{2,1} \right)$} (a12);
        \draw[thick, ->] (a12)  edge node[above left] {$\tilde{\rho}\left( E_{1,2} \right)$} (a7);
        
        \draw[thick, ->] (a10)  edge node[above left] {$\rho\left( E_{1,2} \right),\rho\left( E_{2,3} \right)$} (a9);
        \draw[thick, ->] (a10)  edge node[below right] {$\rho\left( E_{2,1} \right)$} (a11);
        \draw[thick, ->] (a11)  edge node[above left] {$\rho\left( E_{1,2} \right)$} (a10);
        \draw[thick, ->] (a11)  edge node[below right] {$\rho\left( E_{3,2} \right)$} (a12);
        \draw[thick, ->] (a12)  edge node[above left] {$\rho\left( E_{2,3} \right)$} (a11);
        \draw[thick, ->] (a12)  edge node[below right] {$\rho\left( E_{3,2} \right),\rho\left( E_{2,1} \right)$} (a13);
        
        \draw[thick, ->] (a10)  edge node[below] {$\tilde{\rho}\left( E_{2,1} \right)$} (a14);
        \draw[thick, ->] (a11)  edge node[below] {$\tilde{\rho}\left( E_{2,1} \right)$} (a15);
        \draw[thick, ->] (a12)  edge node[below] {$\tilde{\rho}\left( E_{2,1} \right)$} (a16);
\end{tikzpicture}
}%
\end{center}
The vertical columns form identical copies of fundamental $\mathfrak{sl}_3$ representations, while 
horizontal lines form copies of dual $\mathfrak{gl}_2$ representations. Root operators of dual algebra $\mathfrak{gl}_2$ act between different copies of $\mathfrak{sl}_3$ representations. Equations \eqref{eq:27} distinguish the subspace $\text{Span} \left( x_{1}^{1}, x_{2}^{1}, x_{3}^{1} \right)$ as the space of highest vectors of the dual algebra $\mathfrak{gl}_2$:
\begin{align}
\begin{aligned}
    \tilde{\rho}\left( E_{1,2}\right) \, \ket{\lambda} &= 0 \\
    \tilde{\rho}\left( E_{1,1}\right) \, \ket{\lambda}  &= \ket{\lambda} 
\end{aligned}
\end{align}
for $\ket{\lambda} \in \text{Span} \left( x_{1}^{1}, x_{2}^{1}, x_{3}^{1} \right) $.

At the next step of the algorithm we choose variables that will enter the answer for irreducible representations. The number of such variables is equal to the number of positive roots $| \Delta_{+}^{\mathfrak{sl}_{n+1}}| = \frac{(n+1)n}{2}$. The rule and an example for $n = 3$ is provided in Table \ref{Choosing variables A series}.

\begin{table}[h!]
\begin{doublespace}
    \centering
\begin{footnotesize}
    \begin{tabular}{|c|c|}
    \hline
         & $\text{A}_n$ \\
    \hline
        $\begin{matrix}
         &  \\
        \delta_{i}^{a} =  \ x^{a}_{i}, & 1 \leqslant i \leqslant a \leqslant n \\
        \textcolor{red}{v:} \ x^{a}_{i}, & 1 \leqslant a < i \leqslant n + 1 \\
        &  \\
        \end{matrix}$
        & 
        $\begin{pmatrix}
        1 & 0 & 0 \\
        \textcolor{red}{v} & 1 & 0 \\
        \textcolor{red}{v} & \textcolor{red}{v} & 1 \\ 
        \textcolor{red}{v} & \textcolor{red}{v} & \textcolor{red}{v} 
        \end{pmatrix}$\\
    \hline
        $\begin{matrix}
         &  \\
        \textcolor{blue}{e:}  \ \frac{\partial}{\partial x^{a}_{i}}, & 1 \leqslant i \leqslant a \leqslant n \\
        \textcolor{red}{v:} \ \frac{\partial}{\partial x^{a}_{i}}, & 1 \leqslant a < i \leqslant n + 1 \\
         & \\
        \end{matrix}$
        & 
        $\begin{pmatrix}
        \textcolor{blue}{e} & \textcolor{blue}{e} & \textcolor{blue}{e} \\
        \textcolor{red}{v} & \textcolor{blue}{e} & \textcolor{blue}{e} \\
        \textcolor{red}{v} & \textcolor{red}{v} & \textcolor{blue}{e} \\ 
        \textcolor{red}{v} & \textcolor{red}{v} & \textcolor{red}{v} 
        \end{pmatrix}$
        \\
        \hline
    \end{tabular}
\end{footnotesize}
\end{doublespace}
\caption{\footnotesize{In examples we represent $x_{i}^{a}$ as matrices, where $i$ is a number of a row and $a$ is a number of a column. The letter ($\textcolor{blue}{e}$) means expressed variable/derivative, and ($\textcolor{red}{v}$) means variable/derivative that describes irreducible representations. Some variables (free parameters) are set to $0$ or $1$. }}
\label{Choosing variables A series}
\end{table}

One can see from the Table \ref{Choosing variables A series} that upper-triangular part of derivative matrix $\frac{\partial}{\partial x^{a}_{i}}$ will be expressed through the lower-triangular part. That means the upper-triangular part of the variable matrix $x^{a}_{i}$ becomes the free parameters, so we {\it fix the gauge} by setting
\begin{equation}
  \label{eq:28}
\boxed{\  x_{i}^{a} = \delta_{a,i} \, , \qquad 1 \leqslant i \leqslant a \leqslant n }
\end{equation}
Using the constraints~(\ref{eq:28}) we eliminate $x_{i}^{a}$ for
all $i \leqslant a$ from $\rho \left( E_{i,j} \right)$ and then with the help of~(\ref{eq:27})
we eliminate the derivatives $\frac{\partial}{\partial x_{i}^{a}}$ with
$i \leqslant a$. We are left with operators $\rho \left( E_{i,j} \right)$ written entirely in
terms of $x_{i}^{a}$ with $i>a$.

Let us present this procedure more explicitly. Plugging
Eq.~(\ref{eq:28}) into Eq.~(\ref{eq:27}) we get (we omit $P(x_{i}^{a})$ and consider equations as operator equations)
\begin{align}
  \label{eq:29}
&\text{for}\quad  b=a : &\hspace{10mm} \frac{\partial}{\partial x_{a}^{a}} &=  \tilde \lambda_a - \sum_{i=a+1}^{n+1} x_{i}^{a}
    \frac{\partial}{\partial x_{i}^{a}}\\
&\text{for}  b=a+1 : &\hspace{10mm} \frac{\partial}{\partial x_{a}^{a+1}} &= -\tilde{\lambda}_{a+1} x_{a+1}^{a} + 
    \sum_{i=a+2}^{n+1} (x_{a+1}^{a} x_{i}^{a+1} - x_{i}^{a})
    \frac{\partial}{\partial x_{i}^{a+1}}  \label{eq:31}
\end{align}
where we substituted the $\frac{\partial}{\partial x_{a+1}^{a+1}}$ from (\ref{eq:29}).
These two cases $b=a$ and $b = a+1$ are enough to compute simple root generators.

\bigskip

Using Eqs.~(\ref{eq:29}) and~(\ref{eq:31}) we find the Chevalley
generators of $A_n = \mathfrak{sl}_{n+1}$:
\begin{align}
\label{eq:30}
\begin{aligned}
  \mathbf{e}_i &= \rho\left( E_{i,i+1} \right) = \frac{\partial}{\partial x_{i+1}^{i}} +
  \sum_{a=1}^{i-1} x_{i}^{a} \frac{\partial}{\partial x_{i+1}^{a}},\\
  \mathbf{f}_i &=  \rho\left( E_{i+1,i} \right) = \lambda_i \, x_{i+1}^{i} - \left( x_{i+1}^{i} \right)^2 \frac{\partial}{\partial x_{i+1}^{i}} + 
  \sum_{a=1}^{i-1} x_{i+1}^{a} \frac{\partial}{\partial x_{i}^{a}} -
  \sum_{a=i+2}^{n+1} x_{a}^{i} \frac{\partial}{\partial x_{a}^{i+1}}  +
  x_{i+1}^{i} \sum_{a=i+2}^{n+1} \left( x_{a}^{i+1} \frac{\partial}{\partial
      x_{a}^{i+1}} - x_{a}^{i} \frac{\partial}{\partial
      x_{a}^{i}}   \right),  \\
  \mathbf{h}_i &= \rho\left( E_{i,i} \right) - \rho\left( E_{i+1,i+1} \right) = \lambda_i -2 x_{i+1}^{i} \frac{\partial}{\partial x_{i+1}^{i}} +
  \sum_{a=1}^{i-1} \left(  x_{i}^{a} \frac{\partial}{\partial x_{i}^{a}} -
   x_{i+1}^{a} \frac{\partial}{\partial x_{i+1}^{a}} \right)+ 
  \sum_{a=i+2}^{n+1} \left( x_{a}^{i+1} \frac{\partial}{\partial x_{a}^{i+1}} -  x_{a}^{i} \frac{\partial}{\partial x_{a}^{i}} \right).
\end{aligned}
\end{align}
where $i = 1,\ldots,n$ and $\lambda_{i} = \tilde{\lambda}_{i} - \tilde{\lambda}_{i+1}$ are $\mathfrak{sl}_{n+1}$ weights. In these formulas the summation variable is always $a$, therefore one should be careful since it appears both in the upper and lower indices. After  relabeling of indices
\begin{equation}
  \label{eq:32}
  x_{i}^{a} = X_{n+1-a,i-a}
\end{equation}
the generators~(\ref{eq:30}) coincide with the
generators, found in \cite{MRTZ1}. The generators describe irreducible representations in a {\it universal} way, in a sense that the vector fields are nearly independent of a representation. The representation dependence encoded only in simple items with weights $\lambda_{i}$. These formulas are generalizations of the well-known $\mathfrak{sl}_2$ representation($n = 1$ case of \eqref{eq:30} after substitution $x_{2}^{1} = x$):
\begin{equation}
    \mathbf{e}_1 = \frac{\partial}{\partial x} \hspace{10mm}  \mathbf{f}_1 = \lambda_1 x - x^2 \frac{\partial}{\partial x} \hspace{10mm}  \mathbf{h}_1 = \lambda_1 -2 x \frac{\partial}{\partial x} \hspace{10mm} 
\end{equation}

\subsection{$\text{B}_n, \text{C}_n,\text{D}_n$ series}
\label{sec:BCD}
The construction of polynomial representations starts from the fundamental/defining representation of the classical Lie algebras.
\begin{table}[h]
\begin{footnotesize}
\begin{doublespace}
\centering
    \begin{tabular}{|c|c|c|}
    \hline
    $\text{A}_n$ & $\mathfrak{sl}_{n+1} = \Big\{ X \in \text{Mat}_{(n+1)\times(n+1)}  \Big|  \Tr X = 0, \ \Big\}$ &   \\
    \hline
    \hline
    $\text{B}_n$ & $\mathfrak{so}_{2n+1} = \Big\{ X \in \text{Mat}_{(2n+1)\times(2n+1)}  \Big|  X^{T} g + g X = 0, \ g^T = g \Big\}$ & $X = \begin{pmatrix} A & B & a \\ C & -A^T & b \\ - b^{T} & -a^{T} & 0 \end{pmatrix}, \ B^T = -B, \ C^T = -C$  \\
    \hline
    \hline 
    $\text{C}_n$ & $\mathfrak{sp}_{2n} = \Big\{ X \in \text{Mat}_{(2n)\times(2n)}  \Big|  X^{T} g + g X = 0, \ g^T = -g \Big\}$ & $X = \begin{pmatrix} A & B \\ C & -A^T  \end{pmatrix}, \ B^T = B, \ C^T = C$  \\
    \hline
    \hline
    $\text{D}_n$ & $\mathfrak{so}_{2n} = \Big\{ X \in \text{Mat}_{(2n)\times(2n)}  \Big|  X^{T} g + g X = 0, \ g^T = g \Big\}$ & $X = \begin{pmatrix} A & B \\ C & -A^T  \end{pmatrix}, \ B^T = -B, \ C^T = -C$  \\
    \hline
    \end{tabular}
\end{doublespace}
\end{footnotesize}
\caption{\footnotesize{General information about classical Lie algebras. The last column provides form of the block parameterization that is useful for our analysis. } }
\label{fund repr ABCD}
\end{table}

The commutational relations for $\mathfrak{sp}_{2n}$ ($\epsilon = 1$) and $\mathfrak{so}_{2n}$ ($\epsilon = -1$) can be expressed in a convenient form that reflects block structure from Table \ref{fund repr ABCD}:
\begin{align}
\label{so sp generators}
    \begin{aligned}
      A_{i,j} &:= E_{i,j} - E_{j+n, i+n} \\
      B_{i,j} &:= E_{i,j+n} + \epsilon \, E_{j, i+n} \\
      C_{i,j} &:= E_{i+n,j} + \epsilon \,  E_{j+n, i}
    \end{aligned}
\end{align}
\begin{align}
\label{so(2n) and sp(2n) commutators}
    \begin{aligned}
      \left[ A_{i,j} , A_{k,l} \right] &= \delta_{k,j} \, A_{i,l}  - \delta_{i,l} \, A_{k,j}  &\hspace{10mm} \left[ A_{i,j} , B_{k,l} \right] &= \delta_{k,j} \, B_{i,l}  + \delta_{l,j} \, B_{k,i}  \\
      \left[ B_{i,j} , B_{k,l} \right] &= 0 &\hspace{10mm}  \left[ A_{i,j} , C_{k,l} \right] &= - \delta_{i,l} \, C_{k,j}  - \delta_{i,k} \, C_{j,l}  \\
      \left[ C_{i,j} , C_{k,l} \right] &= 0 &\hspace{10mm} \left[ B_{i,j} , C_{k,l} \right] &= \delta_{k,j} \, A_{i,l}  + \delta_{i,l} \, A_{j,k} + \epsilon \, \delta_{j,l} \, A_{i,k}  + \epsilon \, \delta_{i,k} \, A_{j,l}  \\
    \end{aligned}
\end{align}

Another way to deal with classical Lie algebras is to consider simple root generators $\mathbf{e}_i,\mathbf{f}_i,\mathbf{h}_i, \ i = 1,\ldots, \rank \mathfrak{g}$  in the Chevalley basis:
\begin{align}
    \label{KM_rel}
    \begin{aligned}
         [\textbf{h}_i, \textbf{h}_j] &= 0 &\hspace{10mm}  [\textbf{h}_i, \textbf{e}_j] &= \mathcal{A}_{ji} \textbf{e}_j &\hspace{10mm}  \left[ \text{ad}_{\textbf{e}_i} \right]^{1-\mathcal{A}_{ji}} \textbf{e}_j &= 0 \\
          [\textbf{e}_i, \textbf{f}_j] &= \delta_{ij} \textbf{h}_j &\hspace{10mm}  [\textbf{h}_i, \textbf{f}_j] &= -\mathcal{A}_{ji} \textbf{f}_j &\hspace{10mm}  \left[ \text{ad}_{\textbf{f}_i} \right]^{1-\mathcal{A}_{ji}} \textbf{f}_j &= 0 \\
    \end{aligned}
\end{align}
where $\mathcal{A}_{ij}$ is a Cartan matrix. Note that the transposed Cartan matrix enters the relations. To fix the notation we provide examples of Cartan matrices of rank 3.
\begin{equation}
\begin{small}
    \mathcal{A}_{\text{A}_3} = \begin{pmatrix}
    2 & -1 & 0 \\
    -1 & 2 & -1 \\
    0 & -1 & 2
    \end{pmatrix}
    \hspace{10mm}
    \mathcal{A}_{\text{B}_3} = \begin{pmatrix}
    2 & -1 & 0 \\
    -1 & 2 & -2 \\
    0 & -1 & 2
    \end{pmatrix}
    \hspace{10mm}
    \mathcal{A}_{\text{C}_3} = \begin{pmatrix}
    2 & -1 & 0 \\
    -1 & 2 & -1 \\
    0 & -2 & 2
    \end{pmatrix}
    \hspace{10mm}
    \mathcal{A}_{\text{D}_3} = \begin{pmatrix}
    2 & -1 & -1 \\
    -1 & 2 & 0 \\
    -1 & 0 & 2
    \end{pmatrix}
\end{small}
\end{equation}

\begin{table}[h]
\begin{doublespace}
\centering
    \begin{tabular}{|c|c|c|}
    \hline
    $\text{B}_n$ & $\text{C}_n$ & $\text{D}_n$ \\
    \hline
    \multicolumn{3}{|c|}{$ i = 1,\ldots, n-1 $}\\
    \hline
    \multicolumn{3}{|c|}{$\mathbf{e}_i = E_{i,i+1} - E_{i+1+n,i+n}$} \\
    \multicolumn{3}{|c|}{$\mathbf{f}_i = E_{i+1,i} - E_{i+n,i+1+n}$} \\
    \multicolumn{3}{|c|}{$\mathbf{h}_i = E_{i,i}-E_{i+1,i+1} - E_{i+n,i+n} + E_{i+1+n,i+1+n}$} \\
    \hline
    $\mathbf{e}_n = E_{n,2n+1}-E_{2n+1,2n}$ & $\mathbf{e}_n = E_{n,2n}$ & $\mathbf{e}_n = E_{n-1,2n}-E_{n,2n-1}$ \\
    $\mathbf{f}_n = 2E_{2n+1,n}-2E_{2n,2n+1}$ & $\mathbf{f}_n = E_{2n,n}$ & $\mathbf{f}_n = E_{2n,n-1}-E_{2n-1,n}$ \\
    $\mathbf{h}_n = 2E_{n,n}-2E_{2n,2n}$ & $\mathbf{h}_n = E_{n,n} - E_{2n,2n}$ & $\mathbf{h}_n = E_{n-1,n-1}+E_{n,n} - E_{2n,2n} - E_{2n-1,2n-1}$ \\
    \hline
    \end{tabular}
\end{doublespace}
\caption{\footnotesize{Chevalley basis for classical Lie algebras in terms of elementary matrices $E_{i,j}$.}}
\label{Matrix Chevalley basis ABCD}
\end{table}

We consider the following representation for the elementary matrices $E_{i,j}$:
\begin{equation}
\label{E repr 2}
    \rho\left( E_{i,j} \right) = \sum_{a=1}^{n} x_{i}^{a} \frac{\partial}{\partial x^{a}_{j}}.
\end{equation}
This representation acts on the space of polynomials $\mathbb{C}[x_{i}^{a}]$, where $a = 1, \ldots, n$ and $i = 1, \dots, N$. The number of variables is equal to $N \times n$, where $n$ is a rank of an algebra and $N$ is the size of the defining/fundamental representation (see Table \ref{Dimensions of fundamental rep}).

\begin{table}[h!]
\begin{doublespace}
\centering
    \begin{tabular}{|c|c|c|}
    \hline
     $\text{B}_n$ & $\text{C}_n$ & $\text{D}_n$ \\
    \hline
     \multicolumn{3}{|c|}{$ i = 1,\ldots, n-1 $}\\
     \hline
     \multicolumn{3}{|c|}{$\mathbf{e}_i = \sum\limits_{a = 1}^n \left( x^a_i \frac{\partial}{\partial x^{a}_{i+1}} - x^a_{i+1+n} \frac{\partial}{\partial x^{a}_{i+n}} \right) $} \\
     \multicolumn{3}{|c|}{$\mathbf{f}_i = \sum\limits_{a = 1}^n \left(  x^a_{i+1} \frac{\partial}{\partial x^{a}_{i}} - x^a_{i+n} \frac{\partial}{\partial x^{a}_{i+1+n}} \right) $} \\
     \multicolumn{3}{|c|}{$\mathbf{h}_i = \sum\limits_{a = 1}^n \left( x^a_i \frac{\partial}{\partial x^{a}_{i}} -x^a_{i+1} \frac{\partial}{\partial x^{a}_{i+1}} \right)  - \sum\limits_{a = 1}^n  \left( x^a_{i+n} \frac{\partial}{\partial x^{a}_{i+n}} - x^a_{i+1+n} \frac{\partial}{\partial x^{a}_{i+1+n}} \right) $} \\
    \hline
    $\mathbf{e}_n = \sum\limits_{a = 1}^n \left( x^a_{n} \frac{\partial}{\partial x^{a}_{2n+1}} - x^a_{2n+1} \frac{\partial}{\partial x^{a}_{2n}} \right) $ & $\mathbf{e}_n = \sum\limits_{a = 1}^n x^a_{n} \frac{\partial}{\partial x^{a}_{2n}}$ & $\mathbf{e}_n = \sum\limits_{a = 1}^n \left( x^a_{n-1} \frac{\partial}{\partial x^{a}_{2n}} - x^a_{n} \frac{\partial}{\partial x^{a}_{2n-1}} \right) $ \\
    $\mathbf{f}_n = \sum\limits_{a = 1}^n 2\left( x^a_{2n+1} \frac{\partial}{\partial x^{a}_{n}} - x^a_{2n} \frac{\partial}{\partial x^{a}_{2n+1}} \right) $ & $\mathbf{f}_n = \sum\limits_{a = 1}^n x^a_{2n} \frac{\partial}{\partial x^{a}_{n}}$ & $\mathbf{f}_n = \sum\limits_{a = 1}^n \left( x^a_{2n} \frac{\partial}{\partial x^{a}_{n-1}} - x^a_{2n-1} \frac{\partial}{\partial x^{a}_{n}} \right) $ \\
    $\mathbf{h}_n = \sum\limits_{a = 1}^n 2 \left( x^a_{n} \frac{\partial}{\partial x^{a}_{n}} - x^a_{2n} \frac{\partial}{\partial x^{a}_{2n}} \right) $ & $\mathbf{h}_n = \sum\limits_{a = 1}^n \left(  x^a_{n} \frac{\partial}{\partial x^{a}_{n}} - x^a_{2n} \frac{\partial}{\partial x^{a}_{2n}} \right) $ & $\mathbf{h}_n = \sum\limits_{a = 1}^n \left( x^a_{n-1} \frac{\partial}{\partial x^{a}_{n-1}} + x^a_{n} \frac{\partial}{\partial x^{a}_{n}} \right) - $ \\
    & & - $\sum\limits_{a = 1}^n \left( x^a_{2n} \frac{\partial}{\partial x^{a}_{2n}} + x^a_{2n-1} \frac{\partial}{\partial x^{a}_{2n-1}} \right)$ \\
    \hline
    \end{tabular}
\end{doublespace}
\caption{\footnotesize{Chevalley generators for classical Lie algebras in the representation $\rho$. The formulas are obtained by substitution \eqref{E repr} to generators from Table \ref{Chevalley basis ABCD}.}}
\label{Chevalley basis ABCD}
\end{table}

Substituting \eqref{E repr} to the Table \ref{Matrix Chevalley basis ABCD} one obtains the representations for the classical Lie algebras in terms of variables $x_{i}^{a}$ (see Table \ref{Chevalley basis ABCD}).
The representation appears to be reducible and there are a set of operators forming a dual algebra $\tilde{\mathfrak{g}}$ that commutes with the initial representation \eqref{Algebra and dual commute}. The prototypes for the generators of the commuting algebras are the following invariant structures: 
\begin{itemize}
    \item degree operator $d = \sum\limits_{i=1} x_i \frac{\partial}{\partial x_i}$,
    \item invariant form (symmetric for B,D series, anti-symmetric for C series) g = $\sum\limits_{i,j} g_{i,j} \, x_i  x_j$,
    \item  invariant Laplace operator $\Delta$ = $\sum\limits_{i,j} g^{-1}_{i,j} \, \frac{\partial}{\partial x_i} \frac{\partial}{\partial x_j}$.
\end{itemize}
We need to lift them to the case, when the variables $x_i^a$ have an additional index $a$. This converts scalar operators into their tensor analogues: $d \to d^{a,b}, g \to g^{a,b}, \Delta \to \Delta^{a,b}$.\\
Choosing particular form of the invariant form from Table \ref{Invariant forms} we provide explicit form of the generators of the commuting algebras in Table \ref{Commuting algebras generators}.
\begin{table}[h!]
\begin{doublespace}
    \centering
    \begin{tabular}{c|c|c|c}
        & $\text{B}_n$ & $\text{C}_n$ & $\text{D}_n$ \\
         \hline
         $g$ & $\begin{pmatrix}
         0 & 1_{n} & 0 \\
         1_{n} & 0 & 0 \\
         0 & 0 & 1
         \end{pmatrix}$ & $\begin{pmatrix}
         0 & 1_{n} \\
         -1_{n} & 0 \\
         \end{pmatrix}$ & $\begin{pmatrix}
         0 & 1_{n} \\
         1_{n} & 0 \\
         \end{pmatrix}$\\
    \end{tabular}
\end{doublespace}
\caption{\footnotesize{Explicit form of the invariant (anti)-symmetric forms $g$. This particular form is chosen to simplify further calculations and does not mean loss of generality since we consider complexified Lie algebras. }}
\label{Invariant forms}
\end{table}

\begin{table}[h]
\begin{doublespace}
    \centering
    \begin{tabular}{|c|c|c|}
    \hline
    Algebra $\mathfrak{g}$ & Commuting algebra $\tilde{\mathfrak{g}}$ & Generators of $\tilde{\rho} \left( \tilde{\mathfrak{g}} \right)$ \\
    \hline
    $\mathfrak{sl}_{n+1}$ & $\mathfrak{gl}_{n}$ & $\tilde\rho\left(E_{a,b} \right) = \sum\limits_{i = 1}^{n+1} x^a_{i} \frac{\partial}{\partial x^{b}_{i}}$ \\
    \hline
     & & $\tilde\rho\left(A_{a,b} \right) = \sum\limits_{i = 1}^{2n+1} x^a_{i} \frac{\partial}{\partial x^{b}_{i}} + \left(n + \frac{1}{2} \right) \delta_{a,b}$ \\
     $\mathfrak{so}_{2n+1}$ & $\mathfrak{sp}_{2n}$ & $\tilde\rho\left(B_{a,b} \right) = x_{2n+1}^{a} x_{2n+1}^{b} + \sum\limits_{i = 1}^{n} x^a_{i} x^{b}_{i+n} + x^a_{i+n} x^{b}_{i}$ \\
     & & $\tilde\rho\left(C_{a,b} \right) = \frac{\partial}{\partial x_{2n+1}^{a} } \frac{\partial}{\partial x_{2n+1}^{b} } + \sum\limits_{i = 1}^{n} \frac{\partial}{\partial x^a_{i}} \frac{\partial}{\partial x^{b}_{i+n}} + \frac{\partial}{\partial x^a_{i+n}} \frac{\partial}{\partial x^{b}_{i}}$ \\
     \hline
     & & $\tilde\rho\left(A_{a,b} \right) = \sum\limits_{i = 1}^{2n} x^a_{i} \frac{\partial}{\partial x^{b}_{i}} + \left(n \right) \delta_{a,b}$ \\
     $\mathfrak{sp}_{2n}$ & $\mathfrak{so}_{2n}$ & $\tilde\rho\left(B_{a,b} \right) = \sum\limits_{i = 1}^{n} x^a_{i} x^{b}_{i+n} - x^a_{i+n} x^{b}_{i}$ \\
     & & $\tilde\rho\left(C_{a,b} \right) =  \sum\limits_{i = 1}^{n} \frac{\partial}{\partial x^a_{i}} \frac{\partial}{\partial x^{b}_{i+n}} - \frac{\partial}{\partial x^a_{i+n}} \frac{\partial}{\partial x^{b}_{i}}$ \\
     \hline
     & & $\tilde\rho\left(A_{a,b} \right) = \sum\limits_{i = 1}^{2n} x^a_{i} \frac{\partial}{\partial x^{b}_{i}} + \left(n \right) \delta_{a,b}$ \\
     $\mathfrak{so}_{2n}$ & $\mathfrak{sp}_{2n}$ & $\tilde\rho\left(B_{a,b} \right) = \sum\limits_{i = 1}^{n} x^a_{i} x^{b}_{i+n} + x^a_{i+n} x^{b}_{i}$ \\
     & & $\tilde\rho\left(C_{a,b} \right) =  \sum\limits_{i = 1}^{n} \frac{\partial}{\partial x^a_{i}} \frac{\partial}{\partial x^{b}_{i+n}} + \frac{\partial}{\partial x^a_{i+n}} \frac{\partial}{\partial x^{b}_{i}}$ \\
     \hline
    \end{tabular}
\end{doublespace}
\caption{\footnotesize{The first and second column represents pairs of $\mathfrak{g}$ and $\tilde{\mathfrak{g}}$ algebras. The last column provides explicit form of generators \eqref{so sp generators} in representation $\tilde\rho$ forming commuting algebra $\tilde{\mathfrak{g}}$ and satisfying \eqref{Algebra and dual commute}. Note that the algebra generators $A_{a,b},B_{a,b},C_{a,b}$ are different for $\mathfrak{so}$ and $\mathfrak{sp}$ algebras. One can verify that representations of $\mathfrak{so}_{2n}$, $\mathfrak{sp}_{2n}$ indeed satisfy commutation relations \eqref{so(2n) and sp(2n) commutators}}}
\label{Commuting algebras generators}
\end{table}

\section{Reduction to minimal polynomial representation}
\label{sec:BCD reduction}
\subsection{One more simple example of $\mathfrak{so}_{3}$}

Before going to general considerations we provide a simple example of the reduction procedure. The general idea of the reduction was discussed in the Introduction \ref{sec:intro}. Technically the reduction is similar to the well-known procedure of Hamiltonian reduction, where at the first step the system is localized to the fixed value of the moment map and at the second step the remaining degrees of freedom (free parameters) are dropped out (gauge fixing).
\\
We start with simple root generators in the Chevalley basis \eqref{Chevalley basis ABCD} in terms of three variables $x_{1}^{1}, x_{2}^{1}, x_{3}^{1}$ and corresponding derivatives:

\begin{equation}
      \mathbf{e}_1 = x_{1}^{1} \frac{\partial}{\partial x^1_{3}} - x_{3}^{1} \frac{\partial}{\partial x^1_{2}}, \hspace{10mm}
      \mathbf{f}_1 = 2 x_{3}^{1} \frac{\partial}{\partial x^1_{1}} - 2x_{2}^{1} \frac{\partial}{\partial x^1_{3}}, \hspace{10mm}
      \mathbf{h}_1 = 2 x_{1}^{1} \frac{\partial}{\partial x^1_{1}} - 2x_{2}^{1} \frac{\partial}{\partial x^1_{2}}.
\end{equation}
The Borel subalgebra of the commuting algebra $\mathfrak{sp}_2$ is needed for the reduction:

\begin{align}
\begin{aligned}
    \tilde\rho\left(A_{1,1} \right) &= x_{1}^{1} \frac{\partial}{\partial x^1_{1}} + x_{2}^{1} \frac{\partial}{\partial x^1_{2}} + x_{3}^{1} \frac{\partial}{\partial x^1_{3}} + \frac{3}{2} \\
    \tilde\rho\left(B_{1,1} \right) &= 2 x^{1}_{1} x^{1}_{2} + x^{1}_3 x^{1}_3
\end{aligned}
\end{align}
Following the idea from the Introduction \ref{sec:intro} we write the equations \eqref{Space of highest vectors} that distinguish one particular copy of the $\mathfrak{so}_3$ irreducible representation:

\begin{align}
\begin{aligned}
    \left( x_{1}^{1} \frac{\partial}{\partial x^1_{1}} + x_{2}^{1} \frac{\partial}{\partial x^1_{2}} + x_{3}^{1} \frac{\partial}{\partial x^1_{3}} + \frac{3}{2} \right) P\left(x_{1}^{1}, x_{2}^{1}, x_{3}^{1} \right) &= \tilde\lambda_1 P\left(x_{1}^{1}, x_{2}^{1}, x_{3}^{1} \right) \\
    \left( 2 x^{1}_{1} x^{1}_{2} + x^{1}_3 x^{1}_3 \right) P\left(x_{1}^{1}, x_{2}^{1}, x_{3}^{1} \right) &= 0
\end{aligned}
\end{align}
where $P\left(x_{1}^{1}, x_{2}^{1}, x_{3}^{1} \right)$ is a vector from the $\mathfrak{so}_3$ irreducible representation, that actually is the space of all highest vectors of the dual algebra $\mathfrak{sp}_2$. Further in formulas we omit $P\left(x_{1}^{1}, x_{2}^{1}, x_{3}^{1} \right)$ for simplicity and treating equations as operator equations:
\begin{align}
\begin{aligned}
    x_{1}^{1} \frac{\partial}{\partial x^1_{1}} + x_{2}^{1} \frac{\partial}{\partial x^1_{2}} + x_{3}^{1} \frac{\partial}{\partial x^1_{3}} + \frac{3}{2}  &= \tilde\lambda_1  \\
     2 x^{1}_{1} x^{1}_{2} + x^{1}_3 x^{1}_3  &= 0
\end{aligned}
\end{align}
These equations are analogs of moment maps. We chose the variable $x_{3}^{1}$ and its derivative to be the variable describing the reduced representation. With this choice the remaining variables and derivatives are expressed as follows:
\begin{align}
\label{so3 expressed variables}
    \begin{aligned}
      x_{1}^{1} &= \text{free parameter} &\hspace{20mm} \frac{\partial}{\partial x_{1}^{1}} &= \frac{1}{x^{1}_{1}} \left(\tilde\lambda_1 - x_{3}^1 \frac{\partial}{\partial x^{1}_{3}} \right) \\
      x_{2}^{1} &= -\frac{\left( x_{3}^{1} \right)^2}{2 x_{1}^{1}}  &\hspace{20mm} \frac{\partial}{\partial x_{2}^{1}} &= 0\\
    \end{aligned}
\end{align}
The variable $x_{1}^{1}$ is considered as a free parameter, because corresponding derivative is expressed through another variables. To simplify formulas we set free parameter to constant value
\begin{equation}
\label{so3 gauge fixing}
    x_{1}^{1} = 1
\end{equation}
The derivative with respect $x_{2}^{1}$ is vanished because the variable is expressed thought the other variables. This is a general phenomenon and we discuss one more example further in \ref{Vanished derivatives}. The equation \eqref{so3 gauge fixing} is an analog of a fixing of remaining gauge freedom in the theory of Hamiltonian reduction. With relations \eqref{so3 expressed variables} and \eqref{so3 gauge fixing} the final answer for Chevalley basis of a reduced representation reads (we omit indices of $x_{3}^{1}$ and express $\tilde{\lambda}_1 = \lambda_1 / 2$ via true $\mathfrak{so}_3$  weight) :
\begin{equation}
      \mathbf{e}_1 =  \frac{\partial}{\partial x}, \hspace{10mm}
      \mathbf{f}_1 =  \lambda_1 x - x^2 \frac{\partial}{\partial x } , \hspace{10mm}
      \mathbf{h}_1 = \lambda_1 - 2 x \frac{\partial}{\partial x} .
\end{equation}
These formulas coincide with \eqref{eq:30} for $n=1$ due to relation $\mathfrak{sl}_2 \simeq \mathfrak{so}_3$.
\subsubsection{Remark on the vanished derivatives}
\label{Vanished derivatives}
If one has an algebra $\mathfrak{g}(x_1,\ldots, x_{n}; \partial_1, \ldots, \partial_{n})$ on $n$ variables and an integral of motion $H(x_i)$ such that $[\mathfrak{g}, H] = 0$ one can make a reduction by the following recipe. 
\begin{itemize}
    \item Express (locally) one of the variables by setting the integral of motion to a constant:
\begin{equation}
    H(x_1, \ldots, x_n) = 0 \hspace{5mm} \Rightarrow \hspace{5mm} x_n = h(x_1, \ldots, x_{n-1})
\end{equation}
    \item Set the corresponding derivative to zero:
    \begin{equation}
        \partial_n = 0
    \end{equation}
    \item The reduced algebra on $n-1$ variables has the following form
    \begin{equation}
        \mathfrak{g}_{red} = \mathfrak{g}(x_1, \ldots, x_{n-1}, x_{n} = h(x_1, \ldots, x_{n-1}); \partial_1, \ldots, \partial_{n-1}, \partial_{n} = 0 )
    \end{equation}
\end{itemize}
We argue that this procedure reduces the number of degress of freedom without spoiling the Lie algebra structure, i.e. it is a homomorphism.
Let us demonstrate this fact on the previous example of $\mathfrak{so}_3$. 
\begin{align}
\label{angmom}
    \begin{aligned}
        l_x &= z \partial_y - y \partial_z \\
        l_y &= x \partial_z - z \partial_x \\
        l_z &= y \partial_x - x \partial_y \\
    \end{aligned}
\end{align}
with  "integral of motion": 
\begin{equation}
    H = x^2 + y^2 + z^2
\end{equation}
We express coordinates $x,y,z$ and corresponding derivatives through new variables $a,b,r$:
\begin{align}
    \begin{aligned}
        x &= r a  &\hspace{15mm} \partial_x &=   \frac{1 - a^2}{r}  \partial_{a} -  \frac{ab}{r} \partial_{b} + a \, \partial_{r} \\
        y &= r b &\hspace{15mm} \partial_y &=   -\frac{ab}{r}  \partial_{a} + \frac{1- b^2}{r} \partial_{b} + b \, \partial_{r} \\
        z &= r \sqrt{1- a^2 - b^2}  &\hspace{15mm} \partial_z &=   -\frac{a\sqrt{1- a^2 - b^2}}{r}  \partial_{a} -  \frac{b\sqrt{1- a^2 - b^2}}{r} \partial_{b} + \sqrt{1- a^2 - b^2} \, \partial_{r}
    \end{aligned}
\end{align}
Substituting these formulas to \eqref{angmom} one obtains correct $\mathfrak{so}_3$ representation on two variables $a,b$:
\begin{align}
\label{ang mom ab}
    \begin{aligned}
        l_x &= \sqrt{1- a^2 - b^2} \, \partial_b \\
        l_y &= -\sqrt{1- a^2 - b^2} \, \partial_a \\
        l_z &= b \partial_a - a \partial_b \\
    \end{aligned}
\end{align}
The third variable $r$ is absent in the formulas. Note that \eqref{ang mom ab} can be viewed as the following substitution to the \eqref{angmom}:
\begin{align}
\begin{aligned}
    x &\rightarrow a & \hspace{10mm} \partial_x &\rightarrow \partial_a \\
    y &\rightarrow b & \hspace{10mm} \partial_y &\rightarrow \partial_b \\
    z &\rightarrow \sqrt{1 - a^2 -b^2} & \hspace{10mm} \partial_z &\rightarrow 0 
\end{aligned}
\end{align}

\subsection{Reduction for classical Lie algebras}
The reduction starts with equations of type \eqref{Space of highest vectors}:
\begin{align}
\boxed{
    \begin{aligned}
        \tilde\rho\left( A_{a,b} \right) &= \delta_{a,b} \, \tilde\lambda_a,  \hspace{5mm} a \leqslant b \\
       \tilde\rho\left( B_{a,b} \right) &= 0\\
    \end{aligned}
    }
\end{align}

Note that eigenvalues $\tilde\lambda_a$ are different for each series and $B_{a,b}$ operators are different for $\mathfrak{sp},\mathfrak{so}$ algebras \eqref{so sp generators}. These equations are written in operator form, omitting the vector $P\left( x_{i}^{a} \right)$ from the space of highest vectors. We provide explicit form of these equations for each series in the Table \ref{Reduction constraints}.

\begin{table}[h!]
    \centering
\begin{doublespace}
    \begin{tabular}{|c|c|c|}
        \hline
         $\text{B}_n$ & $\text{C}_n$ & $\text{D}_n$ \\
        \hline
        \multicolumn{3}{|c|}{$a \leqslant b$} \\
        \hline
        $\sum\limits_{i = 1}^{2n+1} x^a_{i} \frac{\partial}{\partial x^{b}_{i}}  =  \delta_{a,b} \left(\tilde\lambda_{a} - n -\frac{1}{2} \right)$  & $\sum\limits_{i = 1}^{2n} x^a_{i} \frac{\partial}{\partial x^{b}_{i}}  =  \delta_{a,b} \left(\tilde\lambda_{a} - n \right)$ & $\sum\limits_{i = 1}^{2n} x^a_{i} \frac{\partial}{\partial x^{b}_{i}}  =  \delta_{a,b} \left(\tilde\lambda_{a} - n \right)$ \\
        \hline
        \multicolumn{3}{|c|}{$\text{for all} \ a,b$} \\
        \hline
        $ \sum\limits_{i = 1}^{n} x^a_{i} x^{b}_{i+n} + x^a_{i+n} x^{b}_{i} +  x_{2n+1}^{a} x_{2n+1}^{b} = 0$ & $\sum\limits_{i = 1}^{n} x^a_{i} x^{b}_{i+n} - x^a_{i+n} x^{b}_{i} = 0$ 
        &$\sum\limits_{i = 1}^{n} x^a_{i} x^{b}_{i+n} + x^a_{i+n} x^{b}_{i} = 0 $ \\
        \hline
    \end{tabular}
\end{doublespace}
\caption{\footnotesize{Equations of type \eqref{Space of highest vectors} describing a particular copy of irreducible representation, that is the space of highest vectors of the dual/commuting algebra. In the language of Hamiltonian reduction these equations fix the value of the moment map.}}
\label{Reduction constraints}
\end{table}

To solve the constraints from Table \ref{Reduction constraints} we choose which variables will be expressed and which ones will be the variables of the reduced representations (see Table \ref{BCD variables}). To clarify this idea we provide examples for each series for $n = 3$.

\begin{table}[h!]
\begin{doublespace}
    \centering
\begin{footnotesize}
    \begin{tabular}{|c|c|c|c|c|}
    \hline
         &  $\text{B}_n$ & $\text{C}_n$ & $\text{D}_n$ \\
    \hline
        $\begin{matrix}
        \delta_{i,a} =  \ x^{a}_{i}, & 1 \leqslant i \leqslant a \leqslant n \\
        \textcolor{red}{v:} \ x^{a}_{i}, & 1 \leqslant a < i \leqslant n \\
        \textcolor{blue}{e:} \, x^{a}_{i+n}, & 1 \leqslant i \ \leqslant(<)^{\text{C}_{n}}  \ a \leqslant n \\
        \textcolor{red}{v:} \ x^{a}_{i + n}, & 1 \leqslant a \ <(\leqslant)^{\text{C}_{n}} \ i \leqslant N - n \\
        \end{matrix}$
        & $\begin{pmatrix}
        1 & 0 & 0 \\
        \textcolor{red}{v} & 1 & 0 \\
        \textcolor{red}{v} & \textcolor{red}{v} & 1 \\ 
        \textcolor{blue}{e} & \textcolor{blue}{e} & \textcolor{blue}{e} \\
        \textcolor{red}{v} & \textcolor{blue}{e} & \textcolor{blue}{e} \\
        \textcolor{red}{v} & \textcolor{red}{v} & \textcolor{blue}{e} \\
        \textcolor{red}{v} & \textcolor{red}{v} & \textcolor{red}{v} \\
        \end{pmatrix}$
        &
        $\begin{pmatrix}
        1 & 0 & 0 \\
        \textcolor{red}{v} & 1 & 0 \\
        \textcolor{red}{v} & \textcolor{red}{v} & 1 \\ 
        \textcolor{red}{v} & \textcolor{blue}{e} & \textcolor{blue}{e} \\
        \textcolor{red}{v} & \textcolor{red}{v} & \textcolor{blue}{e} \\
        \textcolor{red}{v} & \textcolor{red}{v} & \textcolor{red}{v} \\ 
        \end{pmatrix}$
        &
        $\begin{pmatrix}
        1 & 0 & 0 \\
        \textcolor{red}{v} & 1 & 0 \\
        \textcolor{red}{v} & \textcolor{red}{v} & 1 \\ 
        \textcolor{blue}{e} & \textcolor{blue}{e} & \textcolor{blue}{e} \\
        \textcolor{red}{v} & \textcolor{blue}{e} & \textcolor{blue}{e} \\
        \textcolor{red}{v} & \textcolor{red}{v} & \textcolor{blue}{e} \\ 
        \end{pmatrix}$
        \\
    \hline
        $\begin{matrix}
        \textcolor{blue}{e:}  \ \frac{\partial}{\partial x^{a}_{i}}, & 1 \leqslant i \leqslant a \leqslant n \\
        \textcolor{red}{v:} \ \frac{\partial}{\partial x^{a}_{i}}, & 1 \leqslant a < i \leqslant n \\
         0 = \frac{\partial}{\partial x^{a}_{i+n}}, & 1  \leqslant i \ \leqslant(<)^{\text{C}_{n}} \ a \leqslant n \\
        \textcolor{red}{v:} \ \frac{\partial}{\partial x^{a}_{i + n}}, & 1  \leqslant a \ <(\leqslant)^{\text{C}_{n}} \ i \leqslant N - n \\
        \end{matrix}$
        & $\begin{pmatrix}
        \textcolor{blue}{e} & \textcolor{blue}{e} & \textcolor{blue}{e} \\
        \textcolor{red}{v} & \textcolor{blue}{e} & \textcolor{blue}{e} \\
        \textcolor{red}{v} & \textcolor{red}{v} & \textcolor{blue}{e} \\ 
       0 & 0 & 0 \\
        \textcolor{red}{v} & 0 & 0 \\
        \textcolor{red}{v} & \textcolor{red}{v} & 0 \\
        \textcolor{red}{v} & \textcolor{red}{v} & \textcolor{red}{v} \\
        \end{pmatrix}$
        &
        $\begin{pmatrix}
        \textcolor{blue}{e} & \textcolor{blue}{e} & \textcolor{blue}{e} \\
        \textcolor{red}{v} & \textcolor{blue}{e} & \textcolor{blue}{e} \\
        \textcolor{red}{v} & \textcolor{red}{v} & \textcolor{blue}{e} \\ 
        \textcolor{red}{v} & 0 & 0 \\
        \textcolor{red}{v} & \textcolor{red}{v} & 0 \\
        \textcolor{red}{v} & \textcolor{red}{v} & \textcolor{red}{v} \\ 
        \end{pmatrix}$
        &
        $\begin{pmatrix}
        \textcolor{blue}{e} & \textcolor{blue}{e} & \textcolor{blue}{e} \\
        \textcolor{red}{v} & \textcolor{blue}{e} & \textcolor{blue}{e} \\
        \textcolor{red}{v} & \textcolor{red}{v} & \textcolor{blue}{e} \\ 
        0 & 0 & 0 \\
        \textcolor{red}{v} & 0 & 0 \\
        \textcolor{red}{v} & \textcolor{red}{v} & 0 \\ 
        \end{pmatrix}$
        \\
        \hline
        $\dim \left( \textcolor{red}{v} \right) = | \Delta_{+}^{\mathfrak{g}} |_{x} + | \Delta_{+}^{\mathfrak{g}} |_{\frac{\partial}{\partial x}}$ & $2n^2$ & $2n^2$ & $2n^2 - 2n$ \\
        $\dim \left( \textcolor{blue}{e} \right) = | \Delta_{+}^{\tilde{\mathfrak{g}}} | + \rank{\tilde{\mathfrak{g}}}$ & $n^2 + n$ & $n^2$ & $n^2 + n$ \\
        \hline
    \end{tabular}
\end{footnotesize}
\end{doublespace}
\caption{\footnotesize{In examples we represent $x_{i}^{a}$ as matrices, where $i$ is a number of a row and $a$ is a number of a column. The letter ($\textcolor{blue}{e}$) means expressed variable/derivative, and ($\textcolor{red}{v}$) means variable/derivative that describes reduced representation. Some variables (free parameters) are set to $0$ or $1$, i.e. the variables $x_{i}^{a}$ with $1 \leqslant i \leqslant a \leqslant n$ are set to $\delta_{i,a}$. Some derivatives are set to $0$, because the corresponding variable is expressed. Note that some inequalities are valid for $\text{B}_n,\text{D}_n$ series while another is valid for $\text{C}_n$ series. }}
\label{BCD variables}
\end{table}
Our aim is to express simple roots generators in terms of reduces variables. For this purpose it is sufficient to solve only equations \eqref{Reduction constraints} where $b = a$ and  $b = a + 1$. From these equations we get expressions (note that we already set free parapeters to constant values $1$ and $0$) provided in Table \ref{Solution to constraints}.
\begin{table}[h!]
\begin{doublespace}
    \centering
\begin{footnotesize}
    \begin{tabular}{|c|c|c|}
    \hline
     & \text{Equation} & \text{Solution} \\
    \hline
    $\text{B}_n$ & $\tilde\rho\left(B_{a,a}\right) = 0$ & $x_{a+n}^{a} = - \frac{1}{2} \left( x_{2n+1}^{a}\right)^2 - \sum\limits_{i = a+1}^{n} x_{i}^{a} x_{i+n}^{a}$ \\
    & $\tilde\rho\left(A_{a,a}\right) = \tilde\lambda_1$ & $\frac{\partial}{\partial x_{a}^{a}} = \left( \tilde{\lambda}_{a} -  n - \frac{1}{2} \right) - \sum\limits_{i = a+1}^{n} x_{i}^{a} \frac{\partial}{\partial x_{i}^{a}} - \sum\limits_{i = a+1}^{n+1} x_{i+n}^{a} \frac{\partial}{\partial x_{i+n}^{a}}$\\\
    & $\tilde\rho\left(A_{a,a+1}\right) = 0$ & $\frac{\partial}{\partial x_{a}^{a+1}} = - x_{a+1}^{a} \left( \tilde{\lambda}_{a+1} -n -\frac{1}{2} \right) + \sum\limits_{i = a+2}^{n} \left( x_{a+1}^{a} x_{i}^{a+1} - x_{i}^{a} \right) \frac{\partial}{\partial x_{i}^{a+1}} +  $\\
    & & $+ \sum\limits_{i = a+2}^{n+1} \left(  x^{a}_{a+1} x_{i+n}^{a+1} - x_{i+n}^{a}  \right)\frac{\partial}{\partial x_{i+n}^{a+1}}$ \\
    \hline
    \hline
    $\text{C}_n$ & $\tilde\rho\left(B_{a,a+1}\right) = 0$ & $x_{a+n}^{a+1} = x_{a+1+n}^{a} - x_{a+1}^{a} x_{a+1+n}^{a+1} + \sum\limits_{i=a+2}^{n} \left( x_{i}^{a+1} x_{i+n}^{a} - x_{i}^{a} x_{i+n}^{a+1} \right)$\\
    & $\tilde\rho\left(A_{a,a}\right) = \tilde\lambda_1$ & $\frac{\partial}{\partial x_{a}^{a}} = \left( \tilde{\lambda}_{a} - n\right) -\sum\limits_{i=a+1}^{n} x_{i}^{a} \frac{\partial}{\partial x_{i}^{a}} -\sum\limits_{i=a}^{n} x_{i+n}^{a} \frac{\partial}{\partial x_{i+n}^{a}} $\\
    & $\tilde\rho\left(A_{a,a+1}\right) = 0$ & $\frac{\partial}{\partial x_{a}^{a+1}} = -x_{a+1}^{a} \left( \tilde{\lambda}_{a+1} - n \right) + \sum\limits_{i=a+2}^{n} \left( x_{a+1}^{a} x_{i}^{a+1} - x_{i}^{a} \right)\frac{\partial}{\partial x_{i}^{a+1}} +  $\\
    & & $+  \sum\limits_{i=a+1}^{n} \left(  x_{a+1}^{a} x_{i+n}^{a+1} - x_{i+n}^{a} \right) \frac{\partial}{\partial x_{i+n}^{a+1}}$ \\
    \hline
    \hline
    $\text{D}_n$ & $\tilde\rho\left(B_{a,a}\right) = 0$ & $x_{a+n}^{a} =  - \sum\limits_{i = a+1}^{n} x_{i}^{a} x_{i+n}^{a}$ \\
    & $\tilde\rho\left(B_{a,a+1}\right) = 0$ & $x_{a+n}^{a+1} = - x_{a + 1 + n}^{a} + \sum\limits_{i = a+2}^{n} \left( x^{a}_{a+1} x_{i}^{a+1} x_{i+n}^{a+1} - x_{i}^{a+1} x_{i+n}^{a} - x_{i}^{a} x_{i+n}^{a+1} \right)  $\\
    & $\tilde\rho\left(A_{a,a}\right) = \tilde\lambda_1$ & $\frac{\partial}{\partial x_{a}^{a}} = \left( \tilde{\lambda}_{a} -  n \right) - \sum\limits_{i = a+1}^{n} \left(x_{i}^{a} \frac{\partial}{\partial x_{i}^{a}} +  x_{i+n}^{a} \frac{\partial}{\partial x_{i+n}^{a}} \right)$\\
    & $\tilde\rho\left(A_{a,a+1}\right) = 0$ & $\frac{\partial}{\partial x_{a}^{a+1}} = - x_{a+1}^{a} \left( \tilde{\lambda}_{a+1} -n  \right) + \sum\limits_{i = a+2}^{n} \left( x_{a+1}^{a} x_{i}^{a+1}  - x_{i}^{a} \right) \frac{\partial}{\partial x_{i}^{a+1}} + $\\
    & & $ + \sum\limits_{i = a+2}^{n} \left(  x^{a}_{a+1} x_{i+n}^{a+1}  - x_{i+n}^{a} \right)\frac{\partial}{\partial x_{i+n}^{a+1}} $\\
    \hline
    \end{tabular}
\end{footnotesize}
\end{doublespace}
\caption{\footnotesize{Solutions to the system of equations from Table \ref{Reduction constraints}. The equations are solved in the order presented in the Table.}}
\label{Solution to constraints}
\end{table}

Substituting these expressions to the formulas for Chevalley generators from Table \ref{Chevalley basis ABCD} we obtain the representations of classical Lie algebras \eqref{sl(n+1) result},\eqref{so(2n+1) result}, \eqref{sp(2n) result},\eqref{so(2n) result}.\\
\newpage
\section{Explicit formulas for polynomial representations}
\label{sec: results}
\subsection{$\text{A}_n$ series}
\begin{align}
\label{sl(n+1) result}
\begin{small}
\boxed{
\boxed{
\begin{aligned}
  \mathbf{e}_i &= \rho\left( E_{i,i+1} \right) = \frac{\partial}{\partial x_{i+1}^{i}} +
  \sum_{a=1}^{i-1} x_{i}^{a} \frac{\partial}{\partial x_{i+1}^{a}},\\
  \mathbf{f}_i &=  \rho\left( E_{i+1,i} \right) = \lambda_i x_{i+1}^{i} - \left( x_{i+1}^{i} \right)^2 \frac{\partial}{\partial x_{i+1}^{i}} + 
  \sum_{a=1}^{i-1} x_{i+1}^{a} \frac{\partial}{\partial x_{i}^{a}} -
  \sum_{a=i+2}^{n+1} x_{a}^{i} \frac{\partial}{\partial x_{a}^{i+1}}  +
  x_{i+1}^{i} \sum_{a=i+2}^{n+1} \left( x_{a}^{i+1} \frac{\partial}{\partial
      x_{a}^{i+1}} - x_{a}^{i} \frac{\partial}{\partial
      x_{a}^{i}}   \right),  \\
  \mathbf{h}_i &= \rho\left( E_{i,i} \right) - \rho\left( E_{i+1,i+1} \right) = \lambda_i -2 x_{i+1}^{i} \frac{\partial}{\partial x_{i+1}^{i}} +
  \sum_{a=1}^{i-1} \left(  x_{i}^{a} \frac{\partial}{\partial x_{i}^{a}} -
   x_{i+1}^{a} \frac{\partial}{\partial x_{i+1}^{a}} \right)+ 
  \sum_{a=i+2}^{n+1} \left( x_{a}^{i+1} \frac{\partial}{\partial x_{a}^{i+1}} -  x_{a}^{i} \frac{\partial}{\partial x_{a}^{i}} \right),
\end{aligned}
}}
\end{small}
\end{align}
The true $\mathfrak{sl}_{n+1}$ weights:
\begin{align}
\begin{aligned}
    i &= 1,\ldots, n: &\hspace{10mm} \lambda_i &= \tilde{\lambda}_i - \tilde{\lambda}_{i+1} \\
\end{aligned}
\end{align}

\subsection{$\text{B}_n$ series}
\begin{align}
\label{so(2n+1) result}
\begin{small}
\boxed{
\boxed{
\begin{aligned}
\textbf{e}_i &=\frac{\partial}{\partial x_{i+1}^{i}}+\sum_{a=1}^{i-1}\left(x_{i}^{a}\frac{\partial}{\partial x_{i+1}^{a}}-x_{i+1+n}^{a}\frac{\partial}{\partial x_{i+n}^{a}}\right),\\
\textbf{f}_i &= \lambda_{i} \, x_{i+1}^{i} -\left(  x_{i+1}^{i} \right)^2 \frac{\partial}{\partial x_{i+1}^{i}} +\sum_{a=1}^{i-1}\left(x_{i+1}^{a}\frac{\partial}{\partial x_{i}^{a}}-x_{i+n}^{a}\frac{\partial}{\partial x_{i+1+n}^{a}}\right) +\left( \frac{1}{2} \left( x_{2n+1}^{i} \right)^2+\sum_{a=i+2}^{n}
x_{a}^{i} x_{a+n}^{i}\right)\frac{\partial}{\partial x_{i+1+n}^{i}} - \\
&-\sum_{a=i+2}^{n} x_{i+1}^{i} x_{a}^{i}\frac{\partial}{\partial x_{a}^{i}}  - \sum_{a=i+2}^{n+1}  x^{i}_{i+1} x_{a+n}^{i}\frac{\partial}{\partial x_{a+n}^{i}} + \sum_{a=i+2}^{n} \left(x_{i+1}^{i} x_{a}^{i+1} - x_{a}^{i} \right)\frac{\partial}{\partial x_{a}^{i+1}} +\sum_{a=i+2}^{n+1} \left(x_{i+1}^{i} x_{a+n}^{i+1} - x_{a+n}^{i}\right)\frac{\partial}{\partial x_{a+n}^{i+1}} \\
\textbf{h}_i &= \lambda_i - 2x_{i+1}^{i}\frac{\partial}{\partial x_{i+1}^{i}} + \sum_{a=i+2}^{n} \left( x_{a}^{i+1} \frac{\partial}{\partial x_{a}^{i+1}} - x_{a}^{i} \frac{\partial}{\partial x_{a}^{i}} \right) + \sum_{a=i+2}^{n+1} \left( x_{a+n}^{i+1}\frac{\partial}{\partial x_{a+n}^{i+1}} - x_{a+n}^{i}\frac{\partial}{\partial x_{a+n}^{i}}\right)+\\
&+\sum_{a=1}^{i-1}\left(x_{i}^{a} \frac{\partial}{\partial x_{i}^{a}}-x_{i+1}^{a}\frac{\partial}{\partial x_{i+1}^{a}}-x_{i+n}^{a}\frac{\partial}{\partial x_{i+n}^{a}}+x_{i+1+n}^{a}\frac{\partial}{\partial x_{i+1+n}^{a}}\right),\\
\textbf{e}_n &= \frac{\partial}{\partial x_{2n+1}^{n}} + \sum_{a=1}^{n-1} \left(x_{n}^{a}\frac{\partial}{\partial x_{2n+1}^{a}}-x_{2n+1}^{a}\frac{\partial}{\partial x_{2n}^{a}}\right),\\
\textbf{f}_n &=\lambda_n \, x_{2n+1}^{n} - \left( x_{2n+1}^{n} \right)^2 \frac{\partial}{\partial x_{2n+1}^{n}}+\sum_{a=1}^{n-1}2\left(x_{2n+1}^{a} \frac{\partial}{\partial x_{n}^{a}} - x_{2n}^{a} \frac{\partial}{\partial x_{2n+1}^{a}}\right),\\
\textbf{h}_n &= \lambda_n - 2 x_{2n+1}^{n} \frac{\partial}{\partial x_{2n+1}^{n}}+\sum_{a=1}^{n-1} 2 \left( x_{n}^{a} \frac{\partial}{\partial x_{n}^{a}}-x_{2n}^{a}\frac{\partial}{\partial x_{2n}^{a}}\right)
\end{aligned}
}}
\end{small}
\end{align}
The true $\mathfrak{so}_{2n+1}$ weights:
\begin{align}
\begin{aligned}
    i &= 1,\ldots, n-1: &\hspace{10mm} \lambda_i &= \tilde{\lambda}_i - \tilde{\lambda}_{i+1} \\
    i &= n: &\hspace{10mm} \lambda_{n} &= 2 \tilde{\lambda}_{n} - (2n + 1) \\
\end{aligned}
\end{align}

\subsection{$\text{C}_n$ series}
\begin{align}
\label{sp(2n) result}
\begin{small}
\boxed{
\boxed{
\begin{aligned}
\textbf{e}_i &= \frac{\partial}{\partial x_{i+1}^{i}}+\sum_{a=1}^{i-1} x_{i}^{a} \frac{\partial}{\partial x_{i+1}^{a}} - \sum_{a=1}^{i} x_{i+1+n}^{a}\frac{\partial}{\partial x_{i+n}^{a}}  \\
\textbf{f}_i &= \lambda_i \, x_{i+1}^{i} -\left( x_{i+1}^{i}\right)^2 \frac{\partial}{\partial x_{i+1}^{i}} +\sum_{a=1}^{i-1}\left(x_{i+1}^{a} \frac{\partial}{\partial x_{i}^{a}}\right)-\sum_{a=1}^{i}\left(x_{i+n}^{a} \frac{\partial}{\partial x_{i+1+n}^{a}}\right) + \sum_{a=i+1}^n \left( x_{a}^{i} x_{a+n}^{i+1} - x_{a}^{i+1} x_{a+n}^{i} \right) \frac{\partial}{\partial x_{i+1+n}^{i+1}} -\\
&-  \sum_{a=i+2}^n x_{i+1}^{i} x_{a}^{i} \frac{\partial}{\partial x_{a}^{i}} -  \sum_{a=i}^{n} x_{i+1}^{i} x_{a+n}^{i} \frac{\partial}{\partial x_{a+n}^{i}} + \sum_{a=i+2}^n \left( x_{i+1}^{i} x_{a}^{i+1} - x_{a}^{i} \right) \frac{\partial}{\partial x_{a}^{i+1}} 
+  \sum_{a=i+1}^n \left( x_{i+1}^{i} x_{a+n}^{i+1} - x_{a+n}^{i} \right) \frac{\partial}{\partial x_{a+n}^{i+1}}\\
\textbf{h}_i &=\lambda_i - 2x_{i+1}^{i}\frac{\partial}{\partial x_{i+1}^{i}} + \sum_{a=i+2}^{n} \left( x_{a}^{i+1} \frac{\partial}{\partial x_{a}^{i+1}} - x_{a}^{i} \frac{\partial}{\partial x_{a}^{i}} \right) + \sum_{a=i+1}^{n}  x_{a+n}^{i+1}\frac{\partial}{\partial x_{a+n}^{i+1}} -\sum_{a=i}^{n}  x_{a+n}^{i}\frac{\partial}{\partial x_{a+n}^{i}}+ \\
&+\sum_{a=1}^{i-1}\left(x_{i}^{a} \frac{\partial}{\partial x_{i}^{a}}-x_{i+1}^{a}\frac{\partial}{\partial x_{i+1}^{a}} \right) -\sum_{a=1}^{i} x_{i+n}^{a}\frac{\partial}{\partial x_{i+n}^{a}}+ \sum_{a=1}^{i+1} x_{i+1+n}^{a}\frac{\partial}{\partial x_{i+1+n}^{a}},\\
\textbf{e}_n &= \frac{\partial}{\partial x_{2n}^{n}}+\sum_{a=1}^{n-1}x_{n}^{a}\frac{\partial}{\partial x_{2n}^{a}}\\
\textbf{f}_n &= \lambda_n \, x_{2n}^{n}-\left( x_{2n}^{n} \right)^2 \frac{\partial}{\partial x_{2n}^{n}} +\sum_{a=1}^{n-1}x_{2n}^{a} \frac{\partial}{\partial x_{n}^{a}}\\
\textbf{h}_n &= \lambda_n-2x_{2n}^{n} \frac{\partial}{\partial x_{2n}^{n}} +\sum_{a=1}^{n-1}\left(x_{n}^{a}\frac{\partial}{\partial x_{n}^{a}}-x_{2n}^{a}\frac{\partial}{\partial x_{2n}^{a}}\right)
\end{aligned}
}
}
\end{small}
\end{align}
The true $\mathfrak{sp}_{2n}$ weights:
\begin{align}
\begin{aligned}
    i &= 1,\ldots, n-1: &\hspace{10mm} \lambda_i &= \tilde{\lambda}_i - \tilde{\lambda}_{i+1} \\
    i &= n: &\hspace{10mm} \lambda_{n} &=  \tilde{\lambda}_{n} - n \\
\end{aligned}
\end{align}

\subsection{$\text{D}_n$ series}
\begin{align}
\label{so(2n) result}
\begin{small}
\boxed{
\boxed{
\begin{aligned}
  \mathbf{e}_i &= \frac{\partial}{\partial
    x_{i+1}^{i}} + \sum_{a=1}^{i-1} \left( x_{i}^{a}
    \frac{\partial}{\partial x_{i+1}^{a}} - x_{i+n+1}^{a}
    \frac{\partial}{\partial x_{i+n}^{a}} \right),\\
  \mathbf{f}_i &=  \lambda_i \, x_{i+1}^{i} -\left( x_{i+1}^{i} \right)^2 \frac{\partial}{\partial x_{i+1}^{i}}+
  \sum_{a=1}^{i-1} \left( x_{i+1}^{a} \frac{\partial}{\partial x_{i}^{a}}
    - x_{i+n}^{a} \frac{\partial}{\partial x_{i+1+n}^{a}} \right) + \left( \sum_{a=i+2}^n x_{a}^{i} x_{a+n}^{i} \right)
  \frac{\partial}{\partial x_{i+1+n}^{i}} + \\
  &\phantom{=} +  \sum_{a=i+2}^n \left[  x_{i+1}^{i} 
  \left( x_{a}^{i+1} \frac{\partial}{\partial x_{a}^{i+1}} + x_{a+n}^{i+1}
    \frac{\partial}{\partial x_{a+n}^{i+1}}   - x_{a}^{i}
    \frac{\partial}{\partial x_{a}^{i}} - x_{a+n}^{i}
    \frac{\partial}{\partial x_{a+n}^{i}} \right) - 
  x_{a}^{i} \frac{\partial}{\partial x_{a}^{i+1}} - x_{a+n}^{i}
    \frac{\partial}{\partial x_{a+n}^{i+1}}  \right], \\
  \mathbf{h}_i &= 
  \lambda_i -2 x_{i+1}^{i} \frac{\partial}{\partial x_{i+1}^{i}}+ \sum_{a=i+2}^n
  \left( x_{a}^{i+1}
    \frac{\partial}{\partial x_{a}^{i+1}} + x_{a+n}^{i+1}
    \frac{\partial}{\partial x_{a+n}^{i+1}} -x_{a}^{i} \frac{\partial}{\partial x_{a}^{i}} -
    x_{a+n}^{i} \frac{\partial}{\partial x_{a+n}^{i}} \right)+  \\
  &\phantom{=}+ \sum_{a=1}^{i-1} \left( x_{i}^{a}  \frac{\partial}{\partial
      x_{i}^{a}}  - x_{i+1}^{a} \frac{\partial}{\partial x_{i+1}^{a}} - x_{i+n}^{a} \frac{\partial}{\partial x_{i+n}^{a}} + x_{i+1+n}^{a}
    \frac{\partial}{\partial x_{i+1+n}^{a}}\right),\\
    \mathbf{e}_n &=  \frac{\partial}{\partial
    x_{2n}^{n-1}} + \sum_{a=1}^{n-2} \left( x_{n-1}^{a} \frac{\partial}{\partial
    x_{2n}^{a}} -  x_{n}^{a}
  \frac{\partial}{\partial x_{2n-1}^{a}} \right),\\
  \mathbf{f}_n &=  \lambda_n \, x_{2n}^{n-1}-
  \left(x_{2n}^{n-1} \right)^2 \frac{\partial}{\partial x_{2n}^{n-1}} +\sum_{a=1}^{n-2}
  \left( x_{2n}^{a} \frac{\partial}{\partial x_{n-1}^{a}} - x_{2n-1}^{a}
    \frac{\partial}{\partial x_{n}^{a}}\right),\\
  \mathbf{h}_n &=
  \lambda_n -2x_{2n}^{n-1} \frac{\partial}{\partial x_{2n}^{n-1}}  + \sum_{a=1}^{n-2} \left( x_{n}^{a} \frac{\partial}{\partial
      x_{n}^{a}} - x_{2n}^{a} \frac{\partial}{\partial
      x_{2n}^{a}} + x_{n-1}^{a} \frac{\partial}{\partial
      x_{n-1}^{a}} - x_{2n-1}^{a} \frac{\partial}{\partial
      x_{2n-1}^{a}} \right)
\end{aligned}
}}
\end{small}
\end{align}
The true $\mathfrak{so}_{2n}$ weights:
\begin{align}
\begin{aligned}
    i &= 1,\ldots, n-1: &\hspace{10mm} \lambda_i &= \tilde{\lambda}_i - \tilde{\lambda}_{i+1} \\
    i &= n: &\hspace{10mm} \lambda_{n} &=  \tilde{\lambda}_{n} + \tilde{\lambda}_{n-1} - 2n \\
\end{aligned}
\end{align}

\section{Flag varieties}
{\bf Notation:}
Let $G$ be a complex simple Lie group associated with the complex Lie
algebra $\mathfrak{g}$ and let $B$ be the Borel subgroup of $G$. One
can think of $G$ as a group generated by exponents $e^{t
  \mathbf{e}_i}$, $e^{ t \mathbf{f}_i}$ and $e^{ t \mathbf{h}_i}$ of
the generators of $\mathfrak{g}$, then the Borel subgroup is generated
by $e^{t \mathbf{e}_i}$ and $e^{ t \mathbf{h}_i}$ only. The quotient
$G/B$ is the so-called generalized flag variety, a complex manifold on
which $\mathfrak{g}$ acts with vector fields (or first order
differential operators). The highest weight representation
$V_{\lambda}$ of $\mathfrak{g}$ can be identified with the vector
space of global holomorphic sections of a certain complex line bundle
$\mathcal{O}_{\lambda}$ on $G/B$. This bundle is built as
follows. Consider the direct product $ G \times \mathbb{C}$ and let
the generators $e^{t \mathbf{e}_i}$ and $e^{ t \mathbf{h}_i}$ of $B$
act on $ (g,v)\in  G \times \mathbb{C}$ as
\begin{equation}
  \label{eq:24}
  e^{t \mathbf{e}_i} (g, v) = (e^{t \mathbf{e}_i} g, v), \qquad e^{t
    \mathbf{h}_i} (g,v) =  ( e^{t
    \mathbf{h}_i}g, e^{t
    \lambda_i} v),
\end{equation}
where $\lambda_i$ are the components of the highest weight of
$V_{\lambda}$. The quotient of $G \times \mathbb{C}$ by the
action~(\ref{eq:24}) of~$B$ is~$\mathcal{O}_{\lambda}$.
\\

The simplest example of this construction is a vector bundle
$\mathcal{O}(k)$ over $SL(2,\mathbb{C})/B = \mathbb{P}^1$. The
manifold $\mathbb{P}^1$ has homogeneous coordinates $[X : Y]$ which
transform in the defining representation of $SL(2,\mathbb{C})$. The
line bundle $\mathcal{O}(k)$ is the space of degree $k$ homogeneous
polynomials in $X$, $Y$. The total number of such polynomials is
$(k+1)$, which matches the dimension of the spin-$\frac{k}{2}$ representation $V_k$ of
$A_1 = \mathfrak{sl}_2$. The action of the generators of
$\mathfrak{sl}_2$ is given by the following formulas:
\begin{equation}
    \begin{aligned}
         \textbf{e}&=Y \frac{\partial}{\partial X}\\
         \textbf{h}&=Y\frac{\partial}{\partial Y}-X\frac{\partial}{\partial X}\\
         \textbf{f}&=X\frac{\partial}{\partial Y}
    \end{aligned}
    \label{xy}
\end{equation}
Since the polynomials are
homogeneous one can eliminate one of the variables and consider them
as functions of the local coordinate $\frac{X}{Y}$ on one of the
patches in $\mathbb{P}^1$.
\\
\subsection{$A_n$ series and complete flags}
In general for $\mathfrak{g} = \mathfrak{sl}_{n+1}$ the manifold $G/B$ is a flag
variety $\mathrm{Fl}(1,2,\ldots,n+1)$ whose points are sequences of
hyperplanes
\begin{equation}
  \label{eq:23}
  \mathrm{Fl}(1,2,\ldots,n+1) = \{\mathbb{C} \subset \mathbb{C}^2
  \subset \cdots \subset \mathbb{C}^{n+1} \}.
\end{equation}

To describe the sequence of hyperplanes we introduce a basis
$\vec{e}^{(a)}$ in $\mathbb{C}^{n+1}$, such that the one dimensional
hyperplane in Eq.~(\ref{eq:23}) is spanned by $\vec{e}^{(1)}$, the
two-dimensional one is spanned by $\vec{e}^{(1)}$ and $\vec{e}^{(2)}$
and so on. The nondegenerate matrix of coordinates of the basis
vectors $x_{i}^{a} = (\vec{e}^{(a)})_i$ gives an element of
$G=GL(n+1,\mathbb{C})$. 

However, some choices of the basis describe the
same sequences of hyperplanes: one can take instead of $\vec{e}^{(a)}$
a linear combination of $\vec{e}^{(b)}$ with $b \leqslant a$. This ``gauge
freedom'' describes the action of $B$ on $G$. To put it differently,
$G$ acts transitively on sequences of hyperplanes, while the subgroup
which fixes a given sequence is precisely $B$, so the flag variety is
indeed the symmetric space $G/B$. Thus the coordinates on the flag
manifold are complex numbers $x_{i}^{a}$ with $a,i=1,\ldots,n$ up to the
action of lower triangular matrices from $B$. There are $(n+1)^2 -
\frac{(n+1)(n+2)}{2} = \frac{n(n+1)}{2}$ coordinates in total.

We introduce the bundle $\mathcal{O}_{\lambda}$
by fixing the gauge freedom and reducing
the number of variables to $\frac{n(n+1)}{2}$.
For this purpose we consider the action
of $\mathfrak{b}$, the Lie algebra of $B$.
Then (\ref{eq:27}) describes the space of holomorphic sections of $\mathcal{O}_{\lambda}$ as
invariant subspace in the space of polynomials $P(x)$ in the variables $x_{i}^{a}$. The
generators $\tilde{\rho}\left( {E}_{a,b} \right)$ commute with $E_{i,j}$
(this is  valid for all $a,b = 1,\ldots,n+1$, not just $a \leqslant b$), so the
constrained space~(\ref{eq:27}) is invariant under the action of
$E_{i,j}$.
It is essential that the number of gauge
  fixing conditions matches the number of independent gauge
  transformations. Another condition is transversality, i.e.\ that the
  $\frac{n(n+1)}{2} \times \frac{n(n+1)}{2}$ conditions
  $\left[ \tilde{\rho}\left( {E}_{a,b} \right), x_{i}^{c} \right]$ for $a \leqslant b$,
  $c\geqslant i$ form non-degenerate matrix. One can verify that both conditions are indeed satisfied
  in our case.

\subsection{$D_n$ series and isotropic flags}
The variety $G/B$ corresponding to the algebra $D_n =
\mathfrak{so}_{2n}(\mathbb{C})$ is the \emph{isotropic} flag variety
$\mathrm{OFl}(1,2,\ldots,n,2n)$, which is described as
follows. Suppose that the vector space $\mathbb{C}^{2n}$ is equipped
with a nondegenerate symmetric bilinear form $g$. Without loss of
generality we can write
\begin{equation}
  \label{eq:34}
  g = \left(
    \begin{array}{cc}
      0 & 1_{n\times n}\\
      1_{n \times n} & 0
    \end{array}
\right).
\end{equation}
We are dealing with complex vector spaces, so there is no difference
between $g$ having positive or negative eigenvalues. The isotropic
flag variety $\mathrm{OFl}(1,2,\ldots,n,2n)$ is the space of sequences
of hyperplanes
\begin{equation}
  \label{eq:36}
  \mathrm{OFl}(1,2,\ldots,n,2n) = \{ \mathbb{C} \subset \mathbb{C}^2
  \subset \cdots \subset \mathbb{C}^n \subset \mathbb{C}^{2n}\}
\end{equation}
which are isotropic with respect to $g$, i.e.\ for any vector
$\vec{v}$ in any of the hyperplanes one has
\begin{equation}
  \label{eq:35}
  g(\vec{v},\vec{v}) = 0.
\end{equation}
To explicitly describe the flag
variety~(\ref{eq:36}) we introduce a basis of $n$ nondegenerate basis
vectors $\vec{e}^{(a)} \in \mathbb{C}^{2n}$, $a=1,\ldots,n$, so that
\begin{equation}
  \label{eq:37}
  \mathrm{OFl}(1,2,\ldots,n,2n) = \{ \mathbb{C}\vec{e}^{(1)} \subset
  \mathbb{C}\vec{e}^{(1)} \oplus \mathbb{C}\vec{e}^{(2)}
  \subset \cdots \subset \bigoplus_{a=1}^n\mathbb{C}\vec{e}^{(a)} \subset \mathbb{C}^{2n}\}.
\end{equation}
As in the $A_n$ case, there is a residual gauge symmetry, since one
can make a triangular linear transformation on $\vec{e}^{(a)}$:
\begin{equation}
  \label{eq:39}
  \{\vec{e}^{(a)}\} \sim \left\{ \sum_{b \leq a} c^a_b \vec{e}^{(b)} \right\}.
\end{equation}
There are thus $\frac{n(n+1)}{2}$ independent parameters of gauge
transformations. To satisfy the isotropy condition~(\ref{eq:35}) we
have to impose additional conditions on the basis $\vec{e}^{(a)}$,
\begin{equation}
  \label{eq:38}
  g(\vec{e}^{(a)},\vec{e}^{(b)}) = 0, \qquad a,b=1,\ldots,n.
\end{equation}
notice that Eq.~(\ref{eq:38}) is gauge invariant. Counting the
variables, constraints and gauge parameters we find that
$\mathrm{OFl}(1,2,\ldots,n,2n)$ has dimension $2n^2 - \frac{n(n+1)}{2}
- \frac{n(n+1)}{2} = n(n-1)$, which coincides with the dimension of
$SO(2n,\mathbb{C})/B$. Indeed, the dimension of $SO(2n,\mathbb{C})/B$
is given by the number of positive roots in the $D_n$ root system,
i.e.\ $\frac{1}{2} (\dim SO(2n,\mathbb{C}) - \mathrm{rank} \,
SO(2n,\mathbb{C})) = \frac{1}{2} \left( \frac{2n(2n-1)}{2} - n \right)
= n(n-1)$.

The coordinates on $\mathrm{OFl}(1,2,\ldots,n,2n)$ are the components
of the basis vectors $x_{i}^{a} = (\vec{e}^{(a)})_i$ with $a=1,\ldots,
n$, and $i=1,\ldots, 2n$. Therefore, we are going to obtain the
representations of $\mathfrak{so}_{2n}$ on the space of polynomials in
rectangular-matrix variables
$x_{i}^{a}$. The generators of the algebra are expressed as differential
operators in $x_{i}^{a}$.

\section{Conclusion}

In this paper we continued to reexamine the prototypes of the free-field representations
of \cite{GMMOS} in the case of classical Lie algebras.
This time we used the option to reduce large representations \eqref{E repr}
to irreducibles by using its commutativity with the dual algebras \eqref{Commuting algebras generators}.
While this procedure is directly applicable in the case of $A_n$,
for other series $B_n, C_n, D_n$ of classical Lie groups the simplest choices are slightly different
and we described them in some details.
We avoided direct use of Gauss and similar decompositions, which were reviewed
from this perspective in \cite{GKMMMO}, but are not always easy to find beyond the
classical series.
The main goal is to obtain the practically useful formulas from a relatively general reasoning,
which can allow further generalizations to other cases, including exceptional, super and affine 
algebras, before proceeding further to the currently interesting case of DIM family \cite{Awata2016bdm,Awata2016mxc,Awata2016riz,Awata2017cnz,Awata2017lqa,Awata2018svb,Zenkevich2018fzl,Zenkevich2019ayk,Zenkevich2020ufs},
where generically working formalism is still under construction.

\section*{Acknowledgements}

Our work was partly supported by the grants of the Foundation for the Advancement of Theoretical Physics ``BASIS'' (A.M., M.R., N.T.), by the grant of Leonhard Euler International Mathematical Institute in Saint Petersburg № 075–15–2019–1619 (M.R., N.T.), by the RFBR grant 20-01-00644 (N.T.),  by the joint RFBR grants 21-51-46010-CT\_a (A.M., N.T., Y.Z.), 21-52-52004-MOST (A.M., Y.Z.).

\printbibliography

\end{document}